\documentclass[sigconf]{acmart}

\usepackage{multirow}
\usepackage{algorithm}
\usepackage{algorithmic}
\usepackage{float}
\usepackage{tikz}
\usepackage{pifont}
\usepackage{enumitem}
\usepackage{subfigure}
\usepackage{flushend}

\newcommand{\revise}[1]{\textcolor[rgb]{0,0.00,0.00}{#1}}
 \newtheorem{theorem}{Theorem}
 
 \newtheorem{property}[theorem]{property}
 
 \newtheorem{corollary}[theorem]{Corollary}

\begin{document}

\title{Achieving Super-Linear Speedup across Multi-FPGA for Real-Time DNN Inference}

\author{\large{{
Weiwen Jiang$^{1,2,3}$\quad
Edwin H.-M. Sha$^{1}$\quad
Xinyi Zhang$^{2}$\quad
Lei Yang$^{2}$\quad
Qingfeng Zhuge$^{1}$\quad
Yiyu Shi$^{3}$\quad
Jingtong Hu$^{2}$
}
}\\
\vspace{-5pt}
{
\normalsize
$^{1}$ East China Normal University\quad
$^{2}$ University of Pittsburgh\quad
$^{3}$ University of Notre Dame
\\
jiang.wwen@gmail.com
}\\
\vspace{10pt}
}

\begin{abstract}
Real-time Deep Neural Network (DNN) inference with low-latency requirement has become increasingly important for numerous applications in both cloud computing (e.g., Apple's Siri) and edge computing (e.g., Google/Waymo's driverless car).
FPGA-based DNN accelerators have demonstrated both superior flexibility and performance; in addition, for real-time inference with low batch size, FPGA is expected to achieve further performance improvement.
However, the performance gain from the single-FPGA design is obstructed by the limited on-chip resource.
In this paper, we employ multiple FPGAs to cooperatively run DNNs with the objective of achieving super-linear speed-up against single-FPGA design.
In implementing such systems, we found two barriers that hinder us from achieving the design goal:
(1) the lack of a clear partition scheme for each DNN layer to fully exploit parallelism, and (2) the insufficient bandwidth  between the off-chip memory and the accelerator due to the growing size of DNNs.
To tackle these issues, we propose a general framework, ``Super-LIP'', which can support different kinds of DNNs.
In this paper, we take Convolutional Neural Network (CNN) as a vehicle to illustrate Super-LIP.
We first formulate an accurate system-level model to support the exploration of best partition schemes.
Then, we develop a novel design methodology to effectively alleviate the heavy loads on memory bandwidth by moving traffic from memory bus to inter-FPGA links.
We implement Super-LIP based on ZCU102 FPGA boards.
Results demonstrate that Super-LIP with 2 FPGAs can achieve $3.48\times$ speedup, compared to the state-of-the-art single-FPGA design.
What is more, as the number of FPGAs scales up, the system latency can be further reduced while maintaining high energy efficiency.
\end{abstract}

\acmConference[CODES+ISSS]{}{2019}{New York, NY}
\acmYear{2019}
\copyrightyear{2019}

\maketitle

\setlength{\textfloatsep}{6pt}
\setlength{\floatsep}{6pt}
\setlength{\dbltextfloatsep}{6pt}

\section{Introduction}

Deep Neural Networks (DNNs) have been continuously achieving breakthroughs in many challenging AI domains, such as image recognition \cite{krizhevsky2012imagenet}, object detection \cite{ren2017faster}, and natural language processing \cite{young2017recent}.
More recently, DNNs have been applied to process live data in interactive services.
For instance, a trained DNN can be employed to process the live video stream for traffic surveillance and emergency response \cite{zhang2017live}, and to analyze medical scans to help doctors during surgery \cite{balakrishnan2018unsupervised}.
Thus, the design of systems for real-time DNN inference becomes an imminent challenge, which attracts increasing studies from both industry \cite{chung2018serving,fowers2018configurable} and academia \cite{ren2017faster,balakrishnan2018unsupervised,ding2018universal}.

Out of all leading computation platforms for DNNs, FPGAs stand out due to their flexibility and versatility over ASICs and their efficiency over CPUs and GPUs. In addition, FPGAs are expected to be more suitable for real-time DNN inference \cite{chung2018serving,fowers2018configurable} for the following reasons. First, even though ASICs can achieve better latency and energy efficiency, it is prohibitive to design ASIC for each different application or upgrade the design when needed. Second, the real-time inference has rigorous requirements of guaranteed latency to ensure user experience, reliability, and even safety. To meet the hard deadlines, uncertainties in CPU- and GPU-based designs (caused by accessing caches) force them to apply the worst-case analysis with a safety margin \cite{wilhelm2008worst}, which leads to inferior design. 
\revise{In contrast, designers are able to customize FPGAs to make the accelerators process the deterministic timing characteristics, which can avoid the overhead caused by the safety margin.}
Third, real-time inference commonly needs to process data with low batch size, which renders the batch throughput optimization inefficient.
For instance, Google's TPU \cite{jouppi2017datacenter} requires the batch size of at least 16 for energy efficiency while FPGAs can extract parallelism from individual execution instance to reduce latency for the low or even no batching.

While most FPGA-based DNN acceleration has been focusing on single-FPGA platform \cite{ma2017optimizing,shen2017maximizing,suda2016throughput,zhang2015optimizing}, \revise{the growth of resource requirement in DNNs has far exceeded the growth of the resource integrated into one FPGA.} As a result, the limited on-chip resources will hinder the exploitation of model parallelism from further boosting time performance.
To overcome this challenge, \cite{zhang2016energy} proposed to employ multiple FPGAs, on which DNN layers can be processed in a pipelined fashion.
Pipelining designs can achieve high throughput; however, the latency cannot be reduced. 
With the objective of minimizing latency for real-time AI applications, we exploit the parallelisms in DNN layers and concurrently process each DNN layer across multiple FPGAs.
However, we observe that with the growing size of models and volume of input data, a straightforward partition of DNNs to multiple FPGAs leads the severe performance degradation due to insufficient communication bandwidth.

In this paper, we propose a new framework, namely ``Super-LIP'', to address the performance bottlenecks in DNNs.
To illustrate the framework, we take Convolutional Neural Networks (CNNs) as a vehicle, \revise{since CNNs have large amount of intermediate data and more complicated data reuse patterns than Recurrent Neural Networks (RNNs)}, which results in the acceleration for CNNs on multi-FPGA more challenging.
Given a CNN, we first formulate an accurate performance model to detect performance bottleneck in an early stage for the latter optimization of accelerator design.
Compared with the existing model \cite{zhang2015optimizing}, the proposed one is more accurate since we investigate the fine-grained data accesses/communication patterns.
Based on the accurate model, we identify that the communication bottleneck is commonly at accessing off-chip memory.
We propose a novel design, ``XFER'', to take advantages of the high-bandwidth inter-FPGA links by offloading part of the data traffic from the memory bus to these links.
As a result, performance bottleneck on memory bandwidth can be significantly alleviated.

The main contributions made in this paper are threefold:
\begin{itemize}[noitemsep,topsep=0pt,parsep=0pt,partopsep=0pt]
  \item \textbf{Super-LIP Framework.} We build a framework, Super-LIP, to control the exploration of FPGA-based designs for real-time DNN inference to achieve \textbf{Super-Li}near speedu\textbf{p} across multiple FPGAs. Inside Super-LIP, \revise{we have further made two contributions listed as follows}.
  \item \textbf{Accurate Model.} First, \revise{we formulate an accurate analytic model} to quantify the performance-resource trade-off in terms of data access/communication patterns, which can guide designers to better design DNN accelerator and provide insights on how to partition DNNs onto multiple-FPGAs. 
  \item \textbf{XFER Design.} Second, we propose a novel design, XFER, to partition and map DNNs onto multiple FPGAs to exploit high parallelism in DNN layers. XFER can further alleviate the performance bottleneck on memory bandwidth by transferring part of the traffic to inter-FPGA links. 
\end{itemize}

\begin{figure}[t]
  \centering
  \includegraphics[width=3.3136 in]{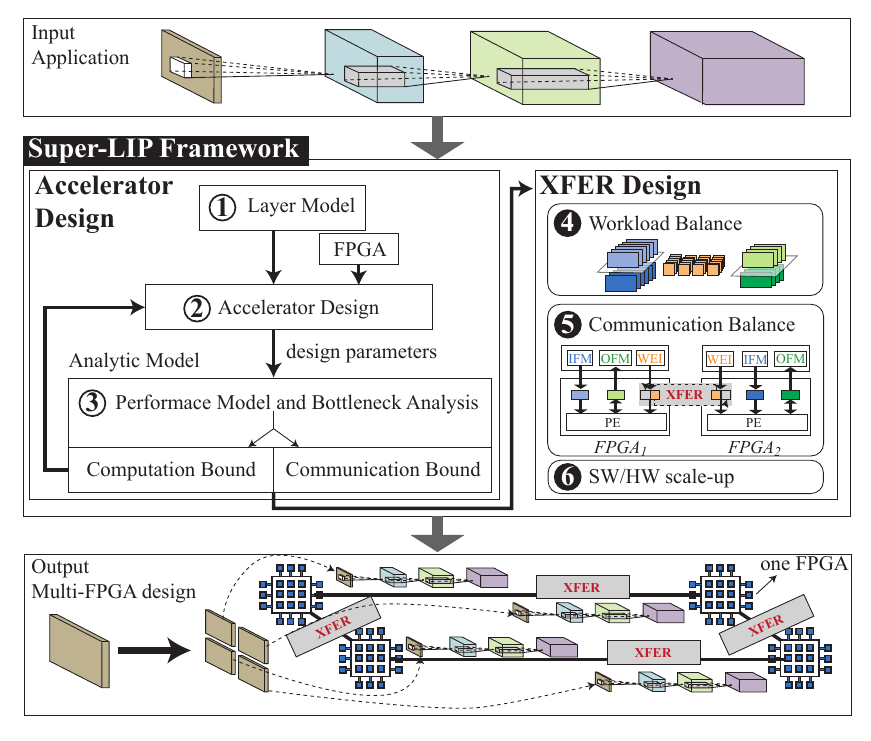}
  \vspace{-20pt}
  \caption{Overview of Super-LIP Framework.}\label{Fig:OVERVIEW}
\end{figure}

Evaluations are conducted on Xilinx ZCU102 FPGA boards connected via optical fiber cables using SFP+ transceiver.
Results show that Super-LIP can achieve $2.82\times$ and $5.81\times$ reduction in latency, while achieving $8.61\times$ and $1.81\times$ improvement in energy efficiency, compared to GPU and mobile GPU, respectively.
Compared with the state-of-the-art single-FPGA design, Super-LIP with 2 FPGAs achieves 3.48$\times$ speedup and 39.86\% improvement in energy efficiency.
In addition, when the size of the FPGA cluster scales up to 16, the latency can be consistently reduced.
For instance, the latency of YOLO \cite{redmon2016you} on one FPGA is \emph{126.6ms}, which can be reduced to \emph{4.53ms}, achieving $27.93\times$ reduction.
This confirms the applicability and scalability of Super-LIP.

The remainder of the paper is organized as follows. Section 2 presents the Super-LIP framework and design challenges. Section 3 and Section 4 present the accurate analytic model and novel XFER design in Super-LIP.
Experimental results are shown in Section 5.
Section 6 discusses related work.
Finally, concluding remarks are given in Section 7.

\section{Super-LIP Framework \& Challenges}

Figure \ref{Fig:OVERVIEW} demonstrates the overview of the proposed Super-LIP framework.
Super-LIP takes a Deep Neural Network (DNN) as input.
And it outputs a multi-FPGA design, onto which the given DNN is partitioned and mapped through a novel technique proposed in this paper.
The design objective of Super-LIP is to achieve the minimum latency for real-time AI with ultra-low batch size.

Super-LIP, in the middle of Figure \ref{Fig:OVERVIEW}, sequentially explores two design spaces: (1) accelerator design space on the left-hand side (\raisebox{-1pt}{\large\ding{192}}-\raisebox{-1pt}{\large\ding{194}}), which determines the hardware-level parallelism (e.g., how many DSPs to compute multiply-accumulate operations, and how many channels to move data between off-chip and on-chip memory);
(2) multi-FPGA design space on the right-hand side (\raisebox{-1pt}{\large\ding{205}}-\raisebox{-1pt}{\large\ding{207}}), which determines task-level parallelism (e.g., how to partition each DNN layer and how to communicate between FPGAs).
In the exploration of both spaces, there exist several challenges needing to be addressed, which are demonstrated as follows.

\vspace{3pt}
\noindent\emph{\textbf{Challenge 1}: Inaccurate performance models hinder designers to optimize accelerators.}
\vspace{3pt}

\begin{figure}[t]
  \centering
  \includegraphics[width= 3.3 in]{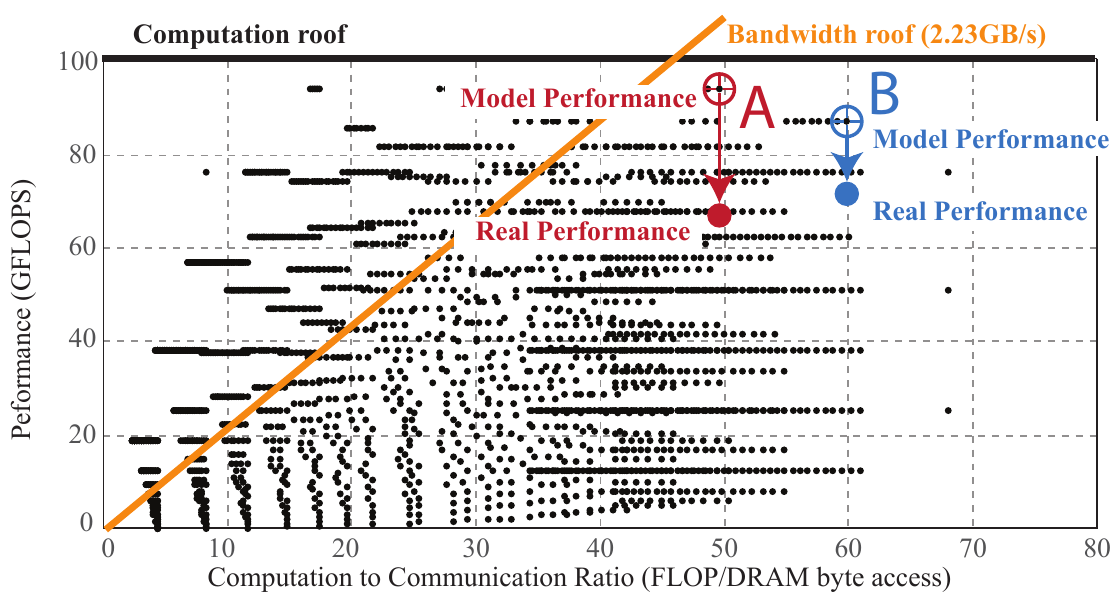}
  \vspace{-10pt}
  \caption{Model performance vs. real performance.}\label{Fig:MotDES}
\end{figure}

Figure \ref{Fig:MotDES} shows the design space exploration of Layer 5 in AlexNet \cite{krizhevsky2012imagenet} using the model proposed by \cite{zhang2015optimizing} which is based on roof-line model.
In this figure, each point represents a design.
Computation roof is determined by the computation resource in FPGA, while bandwidth roof is the theoretic peak memory bandwidth to access off-chip memory in terms of designs.
Design points under these roofs are regarded as attainable.

We implement designs A and B in Figure \ref{Fig:MotDES} on the ZCU102 FPGA board to capture its real performance of on-board execution.
We observe that even though design points A and B are under both roofs, their real performance cannot reach the estimated model performance.
This is because the model in \cite{zhang2015optimizing} assumes the uninterrupted memory access, which is basically impossible due to the synchronization of operations (i.e, the completed operations need to wait for the slower ones, see Figure \ref{Fig:PerfCBD}).
In addition, design A with the best model performance is inferior to design B in real performance.
Therefore, it is imperative to develop an accurate performance model which can explore the optimal designs.

\vspace{3pt}
\noindent\emph{\textbf{Challenge 2}: How to alleviate communication bottleneck without costly modifications on hardware?}
\vspace{3pt}

As shown in Figure \ref{Fig:OVERVIEW} \raisebox{-1pt}{\large\ding{194}}, there are potentially two kinds of performance bottlenecks: computation bottleneck bounded by the computation resource and communication bottleneck bounded by the off-chip memory bandwidth.
For computation bottleneck, we can enlarge the accelerator to involve more computation resource to alleviate it.
However, it is hard to alleviate communication bottlenecks without costly modifications on the FPGA hardware (e.g., \revise{increase the size of on-chip memory}).


Thanks to the inter-FPGA links in an FPGA cluster, memory bandwidth bottleneck can be alleviated by offloading traffic from the memory bus to inter-FPGA links, which can be realized from a high-level implementation without any modifications on hardware.
\revise{This idea is inspired by the observation from the results in comparing data transmission time between accessing off-chip memory and switching between FPGAs. Experimental results on two connected ZCU102 FPGAs via SFP+ cables show that the speed of inter-FPGA communication is competitive with accessing off-chip memory. Specifically, inter-FPGA communication is 3 times faster than accessing off-chip memory when the packet size is 1KB. The figure is 1.6 times when the packet size increases to 64KB and 128KB.
The obtained speedup is mainly because platforms provide high-speed serial communication, while the speed of memory accesses is bounded by the accelerator designs (details in Section \ref{sec:model} \raisebox{-1pt}{\large\ding{193}}-1).}


\begin{figure}
  \centering
  \includegraphics[width=3.3133 in]{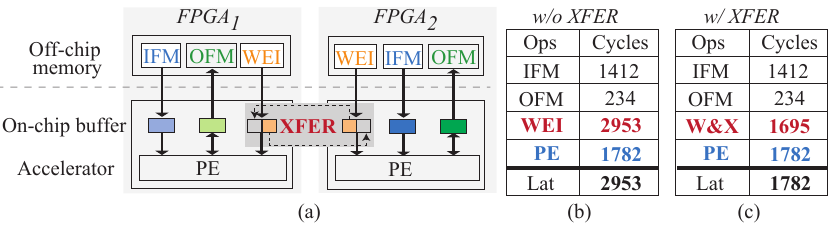}
  \vspace{-20pt}
  \caption{The XFER design and the performance gain.}\label{Fig:MotXFER}
\end{figure}



\revise{Motivated by the above results, we propose a novel design methodology in the Super-LIP framework, namely ``XFER'', to address two implementation problems in exploring multi-FPGA designs}: (1) how to partition computations in DNN layers for higher model parallelism; and (2) what type of data can be off-loaded to inter-FPGA links.
Figure \ref{Fig:OVERVIEW}, from \raisebox{-1pt}{\large\ding{205}} to \raisebox{-1pt}{\large\ding{207}}, demonstrates three steps in XFER to optimize the accelerators on multiple FPGAs: first, it determines the partitions of DNN layers to balance computation workloads (\raisebox{-1pt}{\large\ding{205}}); second, it identifies the traffic to be off-loaded to inter-FPGA links to balance communication (\raisebox{-1pt}{\large\ding{206}}); last, it scales up the number of FPGAs to further speedup the whole DNN network (\raisebox{-1pt}{\large\ding{207}}).


Figure \ref{Fig:MotXFER} (a) illustrates an example of XFER employed between two FPGAs. As shown in the figure, $FPGA_1$ and $FPGA_2$ share the same set of weights and have different sets of input/output feature maps (IFMs/OFMs). Traditionally, each FPGA will load the weight and compute by itself. In XFER, each FPGA only loads half of the shared weight from the off-chip memory. Then, they will send the loaded half weight to each other through inter-FPGA links.
In this way, each FPGA only loads parts of the weights from off-chip memory, which significantly reduces the traffic loads on the memory bus.
\revise{As a result, the overall latency regarding to the pipeline cycle time ($Lat_2$ in Figure \ref{Fig:PerfCBD}) can be reduced from 2,953 to 1,782, achieving 39.65\% improvement, as shown in Figure \ref{Fig:MotXFER} (b)-(c).
Kindly note that the pipeline cycle time is determined by the slowest operation (details can be found in Formulas \ref{equ:lat1} and \ref{equ:lat2}).}

In the following sections, we will address the first challenge by formulating an accurate performance analytic model in Section\ref{sec:model}, which is the base of optimizing DNN accelerators on a multi-FPGA platform.
\revise{Second, in the design of a computing platform with multiple FPGAs where communication channels can be established between two FPGAs, we present the novel XFER design in Section \ref{sec:mfpgad} to address the second challenge.}

\section{Accurate Analytic Models}\label{sec:model}\label{sec:def}

\begin{figure}[t]
  \centering
  \includegraphics[width=2.6 in]{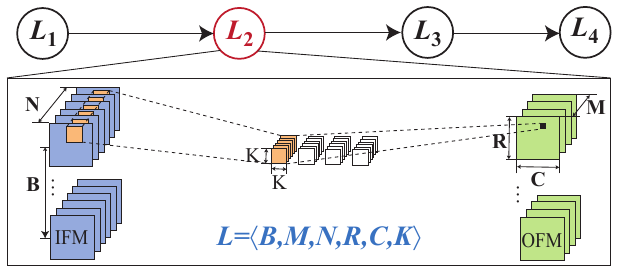}
  \vspace{-10pt}
  \caption{\revise{Super-LIP \large\ding{192}: CNN layer model.}}\label{Fig:LModel}
\end{figure}

\begin{figure}[t]
  \centering
  \includegraphics[width=2.8 in]{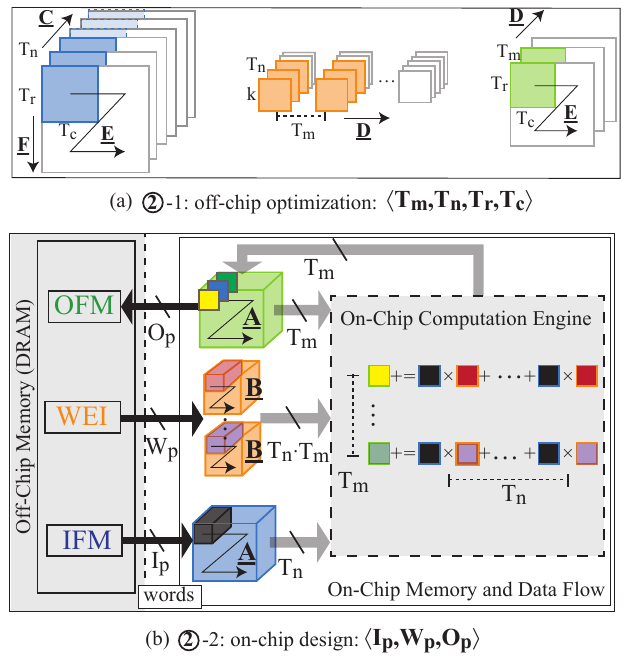}
  \vspace{-10pt}
  \caption{\revise{Super-LIP \large\ding{193}: (a) \large\ding{193}-1 the off-chip optimization; (b) \large\ding{193}-2 the on-chip accelerator design.}}\label{Fig:LDesign}
\end{figure}

\begin{figure}[t]
  \centering
  \includegraphics[width=3.4014 in]{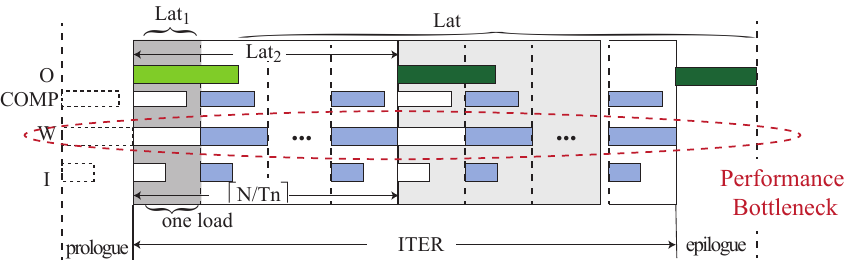}
  \vspace{-10pt}
  \caption{Super-LIP \large\ding{194}: Performance model and bottleneck detection.}\label{Fig:PerfCBD}
\end{figure}

Figures \ref{Fig:LModel}-\ref{Fig:PerfCBD} shows the details of accelerator optimization (left-hand side) in Super-LIP.
We first formulate the model for one CNN layer in \emph{\raisebox{-1pt}{\large\ding{192}}}; then, the accelerator design for both off-chip optimization and on-chip implementations are depicted in \raisebox{-1pt}{\large\ding{193}}-1 and \raisebox{-1pt}{\large\ding{193}}-2, respectively.
At the end of this section, we present the performance model and bottleneck detection (component \raisebox{-1pt}{\large\ding{194}} in Figure \ref{Fig:OVERVIEW}).

\textbf{\emph{\raisebox{-1pt}{\large\ding{192}} Layer Model.}}
\revise{Layer model describes the properties of a CNN layer.
In Figure \ref{Fig:LModel}, we show the details of the second layer $L_2$ in a CNN with 4 layers.}
A CNN layer is defined as $L=\langle B,M,N,R,C,K\rangle$, where $B$ is the batch size; $M$ and $N$ represent the number of channels in output/input feature maps (OFM/IFM); $R$ and $C$ represent the number of rows and columns in OFM; $K$ refers to kernel size.
For example, $L=\langle 2,128,192,13,13,3\rangle$ describes Layer 5 in AlexNet with the batch size of 2.

\revise{Based on the proposed layer model, we will introduce how to design CNN accelerators on an FPGA (component \raisebox{-1pt}{\large\ding{193}} in Figure \ref{Fig:OVERVIEW}).}

\revise{\emph{\textbf{\raisebox{-1pt}{\large\ding{193}} Accelerator Design.}} The core of FPGA-based accelerator design is the on-chip computation engine (as shown in right-hand part of Figure \ref{Fig:LDesign}(b)), which will conduct a set of multiplication-and-accumulation in parallel.
Since the memory and computation requirement of one layer significantly exceeds the on-chip resource, the computation engine cannot process all operations in one CNN layer at once; instead, it will be invoked repeatedly.
To match the speed between the data consumed by the computation engine and the data produced by accessing off-chip memory, we use the on-chip memory (i.e., BRAM) to be a cache, which  constructs a two-level computing model, as shown in Figure \ref{Fig:LDesign}(b).
In the first level, we move data between off-chip memory and on-chip memory, denoted as \raisebox{-1pt}{\large\ding{193}}-1 off-chip design.
In the second level, we move data between on-chip memory and computation engine, denoted as \raisebox{-1pt}{\large\ding{193}}-2 on-chip design.}






\revise{
\textbf{\raisebox{-1pt}{\large\ding{193}}-1 \emph{Off-Chip Design.}} The off-chip design needs to control the sequence of data to be uploaded to the on-chip memory.
To ensure the functional correctness, we need to determine the size of data and the order of data to be uploaded, which correspond to loop tiling and loop ordering.
}

\revise{
A data tile of each data type is the basic unit to be moved between off-chip and on-chip memory.
Let $\langle T_m,T_n,T_r,T_c\rangle$ be tiling parameters on OFM channel, IFM channel, row, column.
Then, we can get the size of data tile for IFM (to be $T_n\cdot T_r\cdot T_c$), OFM (to be $T_m\cdot T_r\cdot T_c$), and weight (to be $T_m\cdot T_n\cdot k\cdot k$, where $k$ is the kernel size).
In Figure \ref{Fig:LDesign}(a), the colored data demonstrate the data tiles. 
Note that these tiling parameters will be constrained by on-chip resource, which will be introduced later in this section.
}

\revise{
Next, the loop order will determine the sequence of data to be moved between off-chip and on-chip memory.
The convolution operation involves 4-level of nested loops (details please refer to Fig. 5 in \cite{zhang2015optimizing}).
These loops traverse along IFM channel, OFM channel, row/column, and batch, which correspond to directions  \textbf{\underline{C}}, \textbf{\underline{D}}, \textbf{\underline{E}}, \textbf{\underline{F}} in Figure \ref{Fig:LDesign}(a), respectively.
}

\revise{
Based on the above loop order, we can get the trip count, which will be used in modeling the computation latency.
We first introduce the trip count for loops along IFM channel (\textbf{\underline{C}}).
According to the tiling parameters, in the loop each step will involve $T_n$ channels, and there are $N$ IFM channels in total.
Therefore, the trip count for direction \textbf{\underline{C}} is $\lceil\frac{N}{T_n}\rceil$.
Similarly, we can obtain the trip count for direction \textbf{\underline{D}} as $\lceil\frac{M}{T_m}\rceil$.
Then, for direction \textbf{\underline{E}}, it will move along $R$ rows and $C$ columns, and each step along row and column is $T_r$ and $T_c$, so the trip count is $\lceil\frac{C}{T_c}\rceil\times \lceil\frac{R}{T_r}\rceil$.
Finally, for batch size of $B$, we need to traverse each batch and the trip count for \textbf{\underline{F}} is $B$.
}

\textbf{\raisebox{-1pt}{\large\ding{193}}-2 \emph{On-Chip Design.}} Based on the loop optimization parameters, we model the on-chip computation and buffer allocation.
Then, we model the off-chip/on-chip communication based on the data width related parameters $\langle I_p,W_p,O_p\rangle$.

For the on-chip computation model, as shown in Figure \ref{Fig:LDesign}(b), there are $T_m\times T_n$ Multiply-Accumulate (MAC) operations conducted in parallel.
In this paper, we consider different data types:
For the 16bits fixed point, each MAC utilizes 1 DSP, while for the 32bits floating point, each MAC utilizes 5 DSPs.
Let $\mathbb{D}$ be the number of DSPs provided by the platform, we have the following constraints for 32bits float-point and 16bits fix-point, respectively:
\begin{equation}\label{equ:dsp_flo}
5\times T_m\times T_n\le \mathbb{D}
\end{equation}
\begin{equation}\label{equ:dsp_fix}
T_m\times T_n\le \mathbb{D}
\end{equation}






As to the on-chip buffer design, there are three kinds of buffers: IFM, OFM and Weight (WEI) buffers.
The size of IFM buffer is determined by the loop tiling on IFM.
For each iteration, $T_n\times T_r\times T_c$ pixels in IFM are loaded to the on-chip buffer, as shown in Figure \ref{Fig:LDesign}(a).
Hence, the IFM buffer is declared as a 3-dimension array $I[T_n][T_r][T_c]$.
Similarly, OFM and WEI buffers are declared as $O[T_m][T_r][T_c]$ and $W[T_m][T_n][K][K]$, respectively.
In order to match the speed of computation, these arrays should be partitioned into different on-chip memories (i.e. BRAM) which can be accessed in parallel.
As shown in Figure \ref{Fig:LDesign}(b), each computation involves $T_n$, $T_m$, and $T_m\times T_n$ pixels/weights in IFM, OFM, and WEI buffers.
Accordingly, we completely partition IFM and OFM along their first dimension, and WEI along its first two dimensions.
Then, we calculate the usage of BRAMs for IFM ($bI$), OFM ($bO$), WEI ($bW$):
\begin{equation}
bI = 2\times T_n\times \lceil T_r\cdot T_c\cdot BITs/18K\rceil
\end{equation}
\begin{equation}
bO = 2\times T_m\times \lceil T_r\cdot T_c\cdot BITs/18K\rceil
\end{equation}
\vspace{-10pt}
\begin{equation}
bW = 2\times T_m\times T_n \times\lceil K\cdot K\cdot BITs/18K\rceil
\end{equation}
where $2$ represents the double-buffer technique adopted in the design.
For the ease of illustration, \revise{we do not put the double buffer in Figure \ref{Fig:LDesign}(b)}.
Let $\mathbb{B}$ be the number of BRAMs in the platform, we have the following constraint.
\begin{equation}\label{equ:bram_all}
bI+bO+bW \le \mathbb{B}
\end{equation}


Finally, we determine the data width parameters $\langle I_p,W_p,O_p\rangle$, which are used to model the off-chip/on-chip communication bandwidth.
$I_p,W_p,O_p$ represent the number of AXI\_STREAMs employed in transmitting IFM, WEI, and OFM, and in turn determine the width of AXI bus.
In a given platform, the data width of the memory bus is limited, denoted as $\mathbb{W}$.
We have the following constraint.
\begin{equation}
BITs\times (I_p+W_p+O_p) \le \mathbb{W}
\end{equation}
where notation $BITs$ is the data bit-width adopted in designs.

\textbf{\emph{\raisebox{-1pt}{\large\ding{194}} Performance Model and Analysis.}}
We first model the off-chip/on-chip communication latency for transferring IFM, OFM, and WEI, based on parameters $\langle I_p,W_p,O_p\rangle$.
For instance, the size of IFM buffer is $T_n\cdot T_r\cdot T_c$, and we employ $I_p$ AXI\_STREAMs
for data transfer, indicating that $I_p$ pixels can be loaded to IFM buffer within 1 clock cycle in the pipelined fashion.
Hence, the latency of loading IFM ($tI_{mem}$) can be formulated.
\begin{equation}\label{equ:tIFM}
tI_{mem} = T_n\cdot T_r\cdot T_c/I_p
\end{equation}
Similarly, we model the latency of loading WEI buffer ($tW_{mem}$) and offloading OFM buffer ($tO_{mem}$) as follows.
\begin{equation}\label{equ:tW}
tW_{mem} = T_m\cdot T_n\cdot K\cdot K/W_p
\end{equation}
\begin{equation}\label{equ:tOFM}
tO_{mem} = T_m\cdot T_r\cdot T_c/O_p
\end{equation}



\begin{figure*}[t]
  \centering
  \includegraphics[width=6.8712 in]{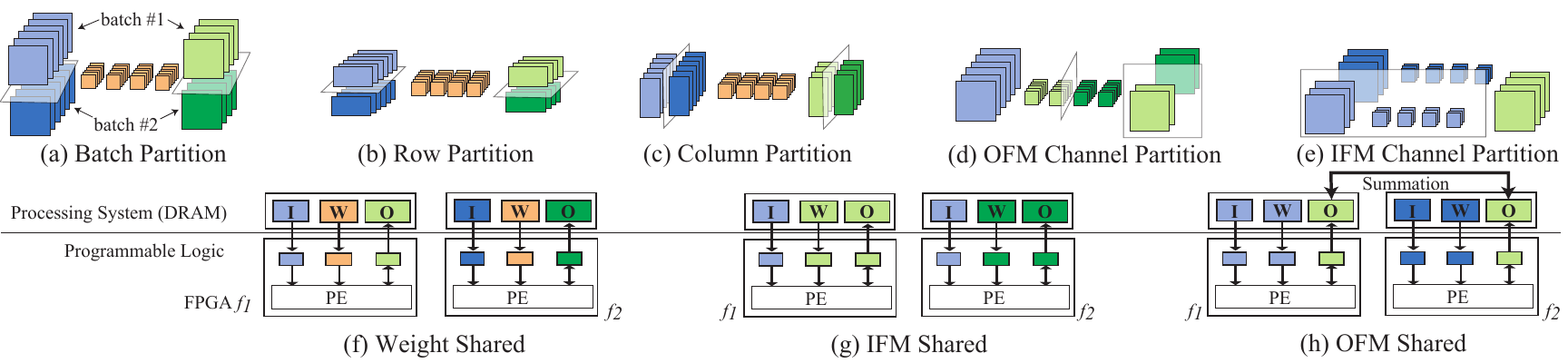}
  \vspace{-12pt}
  \caption{\revise{Five kinds of partitions and the base-line designs: (a) batch partition $P_b=2$; (b) row partition $P_r=2$; (c) column partition $P_c=2$; (d) OFM channel partition $P_m=2$; (e) IFM channel partition $P_n=2$; (f) design for batch/row/column partitions; (g) design for OFM channel partition; (h) design for IFM channel partition.}}\label{Fig:PAR}
\end{figure*}

Then, after filling up the on-chip buffers, data in them can support $K\times K\times T_r\times T_c\times T_m\times T_n$ MACs.
As stated in \raisebox{-1pt}{\large\ding{193}}-2 \emph{On-Chip Design}, the Processing Element (PE) can conduct $T_m\times T_n$ MAC operations in parallel.
Therefore, the latency of one execution of PE ($tComp$) can be modeled as follows.
\begin{equation}
tComp = K\cdot K\cdot Tr\cdot Tc
\end{equation}

Next, we are going to model the system latency.
Benefiting from the double buffer technique, loading IFM, loading WEI, and executing PE can be conducted in parallel.
In addition, loading OFM can be overlapped with $\lceil \frac{N}{T_n}\rceil$ executions, as shown in Figure \ref{Fig:PerfCBD}.
With the known trip counts in \raisebox{-1pt}{\large\ding{193}}-1, we model the latency as follows.
\begin{equation}\label{equ:lat1}
Lat_1 = \max\{tComp,tI_{mem},tW_{mem}\}
\end{equation}
\vspace{-10pt}
\begin{equation}\label{equ:lat2}
Lat_2 = \max\{\lceil \frac{N}{T_n}\rceil\cdot Lat_1,tO_{mem}\}
\end{equation}
\vspace{-6pt}
\begin{equation}\label{equ:Lat}
Lat = B\times \lceil \frac{R}{T_r}\rceil\times \lceil \frac{C}{T_c}\rceil\times \lceil \frac{M}{T_m}\rceil\times Lat_2+ (tO_{mem}+Lat_1)
\end{equation}
where $Lat_1$ and $Lat_2$ are the latencies of one trip for loop \textbf{\underline{C}} and \textbf{\underline{D}}, respectively, and $Lat$ is the overall system latency.

Based on the above performance and resource usage models, we can formulate the optimization problem of minimizing latency as integer non-linear programming problem, which incorporates constraints from Formula \ref{equ:dsp_flo} to \ref{equ:Lat}.
The objective of this problem is to minimize latency, as follows.
\begin{equation}\label{equ:Obj}
Objective: min = Lat
\end{equation}

\textbf{Performance Bottleneck Detection.} The above analytic model can help designer to detect the performance bottleneck. Specifically, we have the following corollary.
\begin{corollary}\label{Corollary1}
Given a CNN layer and the design parameters, we can detect the performance bottlenecks by considering $Lat_1$ and $Lat_2$ as follows:
\begin{itemize}[noitemsep,topsep=0pt,parsep=0pt,partopsep=0pt]
  \item if $Lat_2$ is dominated by $tO_{mem}$, the performance bottleneck is on transmitting OFM data, otherwise,
  \item if $Lat_1$ is dominated by $tI_{mem}$, the performance bottleneck is on transmitting IFM data,
  \item if $Lat_1$ is dominated by $tW_{mem}$, the performance bottleneck is on transmitting weights,
  \item if $Lat_1$ is dominated by $tComp$, we have fully utilized the involved computation resource.
\end{itemize}
\end{corollary}

\section{XFER Design}\label{sec:mfpgad}


In this section, we will present the XFER design for CNNs on multiple FPGAs, which can effectively alleviate the performance bottleneck on off-chip memory bandwidth and in turn achieve the super-linear performance.

\subsection{Design Principles and Objective}

Here, we present three design principles and our ultimate objective for implementing CNNs on multi-FPGA clusters.

$\mathrm{P}1.$ \textbf{Maximizing the utilization of computation resources.}
The design should fully utilize the computation resources. 
In order to achieve this goal, the workloads should be balanced so that FPGAs will not be idle. Meanwhile, we need to avoid memory access stalls, hence DSPs don't have to wait for the data loading from off-chip memory.



%
%


$\mathrm{P}2.$ \textbf{Balancing the traffic loads across multiple FPGAs.}
Different from the single-FPGA design where the memory bus is the only communication channel to access off-chip data, in multi-FPGA clusters, inter-FPGA links provide extra communication bandwidth, which can considerably improve the efficiency of data transfer. However, it is essential to balance the traffic on memory buses and inter-FPGA links for performance improvement.


%

$\mathrm{P}3.$ \textbf{Minimizing the exchange of data in off-chip memories.}
The movements of data stored in off-chip memory are controlled by CPUs, which incurs high latency. Thus, we should avoid such movement as much as possible. 


%

%

$\mathrm{Objective}.$ \textbf{Achieving super-linear speedup.} The design objective is to achieve super-linear performance, such that the latency can be minimized without compromising on throughput or energy efficiency, which leads to the ultimate realization of real-time DNN inference.

Following the above principles and objective, XFER consists of the following three steps:
First, XFER achieves linear speedup through balancing computation workloads.
Then, to achieve further speedup, XFER identifies the shared data among different partitions, and distributes these data across FPGAs to reduce the traffic loads on each FPGA's memory bus.
During run time, FPGAs transfer a part of the data stored in their local off-chip memory via inter-FPGA links.



%



\subsection{Layer Partition and Workload Balance}
\revise{A CNN layer can be partitioned into different parts to be computed in parallel.
The most common partition is the batch partition, where the IFM and OFM are divided along batching direction, as shown in Figure \ref{Fig:PAR}(a).
The computation of a batch of OFM only relies on the corresponding batch of IFM and the whole weights.
In consequence these batches can be computed in parallel in multiple processing elements (PEs) if weights are duplicated to PEs.
Similarly, we can partition CNN layers along rows (R) and columns (C), as shown in Figure \ref{Fig:PAR}(b)-(c).
In addition, we can partition a CNN layer as follows: dividing the OFM into multiple parts along channel direction, and dividing weights correspondingly, as shown in Figure \ref{Fig:PAR}(d).
In this case, we use the whole IFM and part of weights to compute part of OFM, and we call such partition as OFM channel partition.
Similarly, we can partition IFM along channel direction and weights correspondingly, as shown in Figure \ref{Fig:PAR}(e).}

\revise{For each kind of partition, we define a partition factor to indicate the number of parts generated by the partition.
We use notations $P_b,P_r,P_c,P_m,P_n$ to represent factors for partitions along with batch (B), rows (R), columns (C), OFM channels (M), and IFM channels (N).
For instance, $P_r=2$ indicates that we partition IFM/OFM along rows into 2 parts, as shown in Figure \ref{Fig:PAR}(b).
Kindly note that we use factors of 2 in Figure \ref{Fig:PAR} for the simplicity of illustration; however, these factors can be other positive integers except 2, which will be restricted by the number of available FPGAs in a system.}







According to the types of shared data, we can classify partitions into 3 categories:
first, ``weight shared'' case, where the computations of different partitions use the same weights, such as row or column partitions in Figure \ref{Fig:PAR}(a)-(c);
second, ``IFM shared'' case, where the computations under the OFM channel partition use the same IFM, as shown in Figure \ref{Fig:PAR}(d);
third, ``OFM shared'' case, where the computations under the IFM channel partition share the OFM.
Kindly note that ``OFM shared'' will cause the transmission of intermediate data as shown in Figure \ref{Fig:PAR}(h), which violates design principle $\mathrm{P}3$.
Hence, we do not consider it in the designs.

For different types of partitions, we present a straightforward design that follows \emph{\textbf{design principle $\mathrm{P}1$}} to balance workloads among FPGAs.
This design will be the base of XFER.
The main idea of this design is to map each partition to one FPGA and replicate the shared data to each FPGA.
Figure \ref{Fig:PAR}(f) shows the design for weight shared partitions, where IFM and OFM are independent on two FPGAs and the whole weights are replicated to these FPGAs.
Similarly, Figure \ref{Fig:PAR}(g) shows the design of IFM shared partition.

In the above design, since the shared data are replicated to each FPGA, the computations in any two FPGAs have no dependency.
It implies that all FPGAs can be executed in parallel, enabling linear speedup in the DNN inference.

\begin{figure}
  \centering
  \includegraphics[width=3.3307 in]{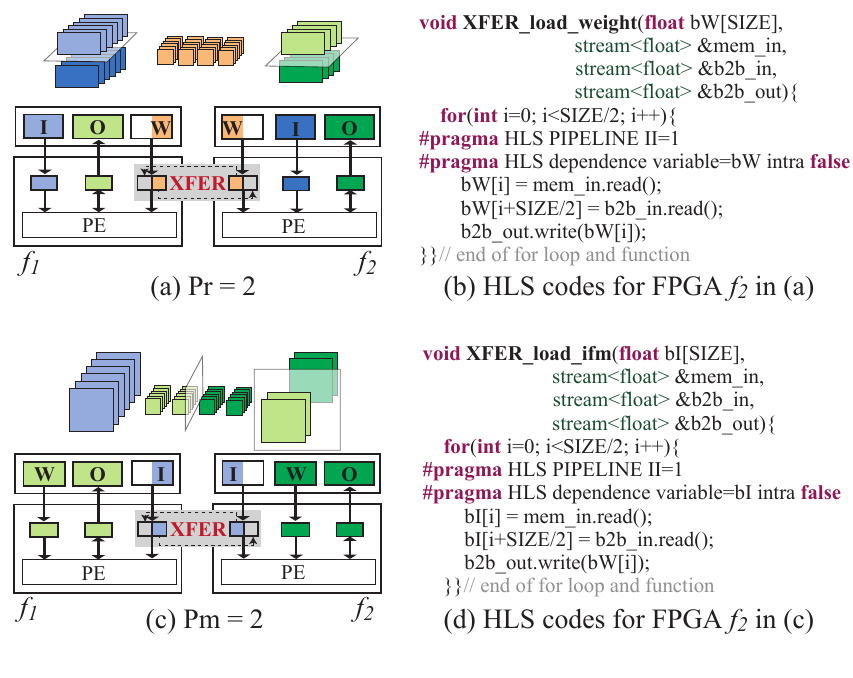}
  \vspace{-12pt}
  \caption{XFER design: (a) weight shared case; (b) HLS codes (a); (c) IFM shared case; (d) HLS codes for (c).}\label{Fig:XFERDes}
\end{figure}

\subsection{Share Data through inter-FPGA links}\label{Sec:XFER}



XFER can further boost the performance by transferring parts of the traffic loads from the memory bus to inter-FPGA links.
Since XFER modifies the data communication subsystem, the matched system model needs to be revised.
In the following text, we will introduce the detailed XFER design for weight shared partition and IFM shared partition.

\textbf{Weight Shared Partition.}
Figure \ref{Fig:XFERDes}(a) demonstrates the XFER design for weight shared partition, where each FPGA load a part of weights from its local off-chip memory, and obtain the remaining parts from its neighbors via inter-FPGA links.
In other words, a copy of the whole weights are distributed in the off-chip memory across FPGAs.
During the execution, each FPGA conducts three operations: (1) loading half of the data from its off-chip memory to on-chip buffer, (2) sending the loaded data to other FPGAs through the inter-FPGA links, and (3) receiving the remaining data from other FPGAs through the inter-FPGA links.



\emph{HLS design.} To efficiently conduct these operations, data are streamed in and out from the on-chip weight buffers (e.g., utilizing the AXI streams for Xilinx FPGAs).
The detailed HLS codes are given in Figure \ref{Fig:XFERDes}(b).
Kindly note that in order to transmit data in parallel, we make the intra dependency on $bW$ to be \emph{false}.
In addition, we use the pipeline pragma to load and send the weight in the pipelined fashion.

\emph{Model modification.}
XFER can reduce the latency of loading weights $tW_{mem}$.
With partition factors $\langle P_b,P_r,P_c\rangle$, $tW_{mem}$ in XFER will be ${P_b\cdot P_r\cdot P_c}$ times smaller.
Thus, we replace Formula \ref{equ:tW} as follows.
\begin{equation}
tW_{mem} = T_m\cdot T_n\cdot K\cdot K/(W_p\cdot P_b \cdot P_r\cdot P_c)
\end{equation}


Then, we will formulate the newly incurred inter-FPGA communications.
In XFER, each FPGA only holds $\frac{1}{P_b\cdot P_r\cdot P_c}$ part of the whole weights.
It requires ${P_b\cdot P_r\cdot P_c}-1$ communication channels.
For each channel, the latency is the same and we model the latency of the $i^{th}$ channel as follows.
\begin{equation}
tW^{i}_{b2b} = T_m\cdot T_n\cdot K\cdot K/(W^{b2b}_p\cdot P_b \cdot P_r\cdot P_c)
\end{equation}
where $W^{b2b}_p$ is the number of ports in one channel.

Finally, the latency $Lat_1$ should be modified, since XFER incurs the inter-FPGA communication in the inner-most loop \textbf{\underline{C}} in Figure \ref{Fig:LDesign}(a).
Formula \ref{equ:lat1} is revised as follows.
\begin{equation}
Lat_1 = max\{tComp,tI_{mem},tW_{mem},\max\limits_{i\in [1,..,P-1]}\{tW_{b2b}^{i}\}\}
\end{equation}
where $P-1$ is the number of communication channels.


\textbf{IFM Shared Partition.}
Similar to the design of the weight shared case, XFER will transfer the movements of IFM from the memory bus to inter-FPGA links, as shown in Figure \ref{Fig:XFERDes}(c)-(d).
We add Formula \ref{equ:newb2b} to model the latency on the $i^{th}$ inter-FPGA link, and replace formulations for modeling the latency of loading IFM and the latency $Lat_1$.
\begin{equation}\label{equ:newb2b}
tI^{i}_{b2b} = T_m\cdot T_n\cdot K\cdot K/(I^{b2b}_p\cdot P_m)
\end{equation}
\begin{equation}
tI_{mem} = T_m\cdot T_n\cdot K\cdot K/(I_p\cdot P_m)
\end{equation}
\begin{equation}
Lat_1 = max\{tComp,tI_{mem},tW_{mem},\max\limits_{i\in [1,..,Q-1]}\{tI_{b2b}^{i}\}\}
\end{equation}
where $I^{b2b}_p$ is the number of data transmitted via inter-FPGA links in parallel and $Q-1=P_m-1$ is the number of links.

\begin{figure}
  \centering
  \includegraphics[width=3.3689 in]{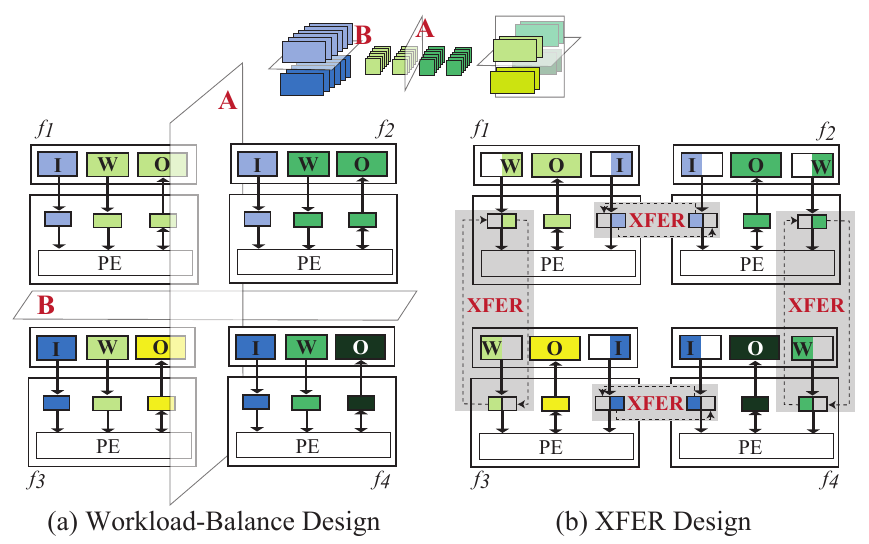}
  \vspace{-12pt}
  \caption{Workload-balance design and XFER design for the partitions with factors $P_r=2$ and $P_m=2$.}\label{Fig:Hybrid}
\end{figure}

%
%
%
%
%
%


\subsection{Extension to Hybrid Partitions}\label{Sec:Hybrid}


We continue to extend XFER to support hybrid partitions where both weights and IFMs are shared.
For partition $\langle P_b, P_r, P_c, P_m\rangle$, XFER involves $P_b\cdot P_r\cdot P_c\cdot P_m$ FPGAs.
In the following texts, we solve three problems in constructing a network of FPGAs: (1) how to organize FPGAs, (2) what's the topology of the network, (3) how to formulate the inter-FPGA bandwidth constraints.

\textbf{Organization.} We organize FPGAs in a two-dimensional array (2D-array) with $P_m$ columns and $P_b\cdot P_r\cdot P_c$ rows by the following two steps.
First, for the IFM shared partition (e.g., partition $A$ in Figure \ref{Fig:Hybrid}), the weights and OFM are partitioned into independent parts and allocated to a column of FPGAs, while the whole IFM is replicated to all FPGAs.
Second, when considering the OFM shared partition (e.g., partition $B$), for all FPGAs in one column, the IFM and OFM are split into independent partitions.
Then, the design with workload balanced can be obtained, as shown in Figure \ref{Fig:Hybrid}(a).
The design has one property as follows.


%
%
%
%
%

\begin{property}
All FPGAs in one column share a part of weights, while all FPGAs in one row share a part of IFMs.
\end{property}

Based on the above property, we extend XFER to support hybrid partitions.
Specifically, for a column of FPGAs, a corresponding part of the weights are distributed among these FPGAs, and exchange the weights during execution like that in the 2-FPGA system.
For a row of FPGAs, they share a part of the IFMs.
Figure \ref{Fig:Hybrid}(b) demonstrates XFER for the hybrid partitions with factors $P_r=2$ and $P_m=2$, where the traffic loads on the memory bus can be significantly reduced.

\begin{figure}
  \centering
  \includegraphics[width=3.3418 in]{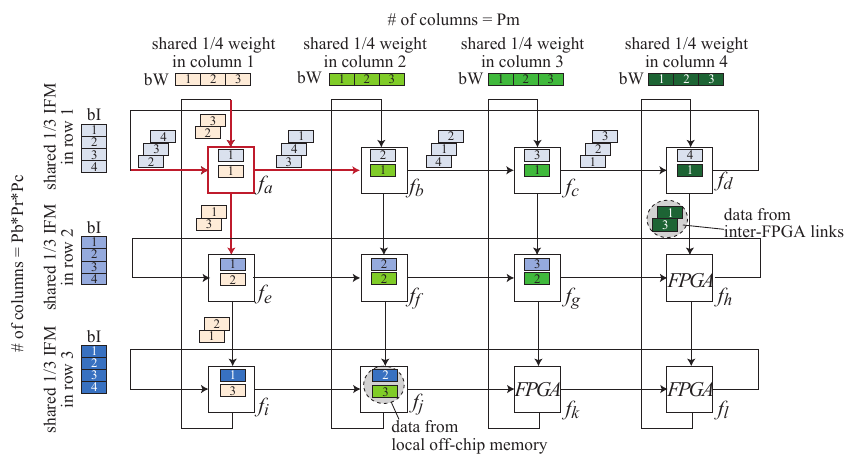}
  \vspace{-12pt}
  \caption{2D-torus topology and data movement.}\label{Fig:Topo}
\end{figure}

\textbf{Topology.} Benefiting from the regular 2D-array organization, we have a wide range of choices for the networking topology, such as mesh, torus, folded torus, etc. 
In this work, we employ the 2D-torus topology to build the connections among FPGAs \cite{yang2017task,yang2016fotonoc}.
One reason to select 2D-torus is that we can apply a uniform design for each FPGA.
As shown in Figure \ref{Fig:Topo}, where $P_m=4$ and $P_b\cdot P_r\cdot P_c=3$, each FPGA has two incoming links and two outgoing links; meanwhile, the amount of data received and transmitted by each FPGA are the same.
In addition, with such a 2D-torus topology, the traffic loads on columns and rows are balanced, which overrides our \emph{\textbf{design principle $\mathrm{P}2$}}.



\textbf{Bandwidth Constraints.} Based on 2D-torus topology, we formulate the bandwidth constraints. 
As shown in Figure \ref{Fig:Topo}, the loads on incoming links and outgoing links are the same; therefore, we build the constraints on one direction.

We first consider the communication on one row for IFM sharing.
The size of shared IFM is bI, which is shared by $P_m$ FPGAs.
Take FPGA $f_a$ in Figure \ref{Fig:Topo} as an example, the total amount of data transmitted on outgoing link will be $D_{row}=(P_m-1)\cdot \frac{bI}{P_m}$.
Similarly, for one column, the total amount of data on outgoing link will be $D_{col}=(P_b\cdot P_r\cdot P_c-1)\cdot\frac{bW}{P_b\cdot P_r\cdot P_c}$.
Kindly note that these data transmissions need to be completed in the time of $Lat_1$ to avoid worsening the whole latency (see Equation \ref{equ:lat1}).
Let $\mathbb{NB}$ be the maximum bandwidth of the inter-FPGA communication on one direction.
We have the following constraint.
\begin{equation}
D_{row}+D_{col}\le \mathbb{NB}\cdot Lat_1
\end{equation}





%
%

\begin{figure}
  \centering
  \includegraphics[width=3.3625 in]{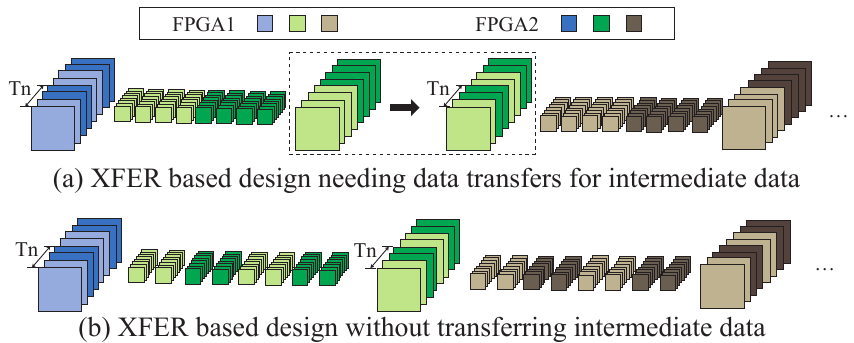}
  \vspace{-12pt}
  \caption{Intermediate data placement across layers.}\label{Fig:MultiLayer}
\end{figure}

\begin{figure*}
  \centering
  \includegraphics[width=7 in]{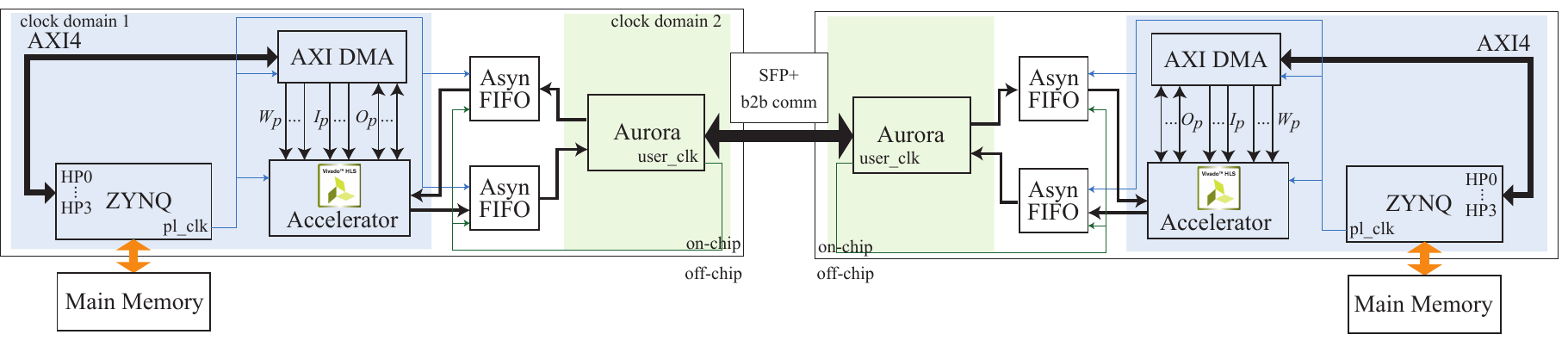}
  \vspace{-20pt}
  \caption{Implementation overview of a multi-FPGA cluster with two FPGAs connected by SFP+.}\label{Fig:ImpOverview}
\end{figure*}

\subsection{Extension to Multiple CNN Layers}\label{Sec:PerfAnaly}

We have discussed the optimizations for single CNN layer on multiple FPGAs.
However, due to CNNs have many layers, how to seamlessly execute multiple layers on multiple FPGAs is important.
This subsection will follow our \emph{\textbf{design principle $\mathrm{P}3$}} to extend XFER to support multiple CNN layers.
To make data remain in-situ to minimize the intermediate data movements, the partition schemes for different layers should be coordinated according to XFER.



Figure \ref{Fig:MultiLayer} shows two cases for the IFM shared partitions.
The partition in Figure \ref{Fig:MultiLayer}(a) will lead the exchange of intermediate data between FPGAs $f_1$ and $f_2$.
Because in XFER (Figure \ref{Fig:XFERDes}(d)), the first of $Tn/2$ channels of IFM are loaded to $f_1$, and the remained $Tn/2$ channels are loaded to $f_2$. This will be applied to all of the channels until the end of the IFM.
With the partition shown in Figure \ref{Fig:MultiLayer}(a), all first $T_n$ channels in OFM are produced at $f_1$.
In order to make it work for the convolution operation in the next layer, we need to exchange half of the OFM between $f_1$ and $f_2$.
In contrast, if the OFM channels are partitioned in an interleaving way as shown in Figure \ref{Fig:MultiLayer}(b), no data movements are required across layers.
%
%

Based on the above observations, we investigate different partition methods for consecutive layers.
For batch partition, the required data for the next layer on an FPGA is totally produced by itself, and no data movement is required.
For row/column partition, only borders need to be transferred, where the small number of data can be transmitted during the execution via inter-FPGA links without passing CPUs.
For OFM channel partition, we can avoid data movements by employing the interleaving partition, as shown in Figure \ref{Fig:MultiLayer}(b).
Furthermore, we found that if the consecutive layers employ different partition methods, data movement is unavoidable.
In consequence, we will deploy CNNs on multi-FPGA clusters with uniform partition factors across CNN layers.

\subsection{Analysis of Performance}\label{Sec:PerfAnaly}

In the following, we analyze the performance in both latency and elapsed time to explore the design space.

\textbf{Single Layer.}
XFER can achieve super-linear speedup due to the following two reasons.
First, the trip counts of loops \textbf{\underline{D}}, \textbf{\underline{E}}, \textbf{\underline{F}} can be linearly reduced.
Second, by alleviating the performance bottleneck on accessing off-chip memory, the latency $Lat_1$ can be further reduced.
In consequence, XFER is able to achieve super-linear speedup.

\begin{table}[t]
  \centering
  \tabcolsep 2.8pt
  \renewcommand\arraystretch{1.3}
  \footnotesize
  \caption{\revise{Layer-specific and cross-layer optimization}}
  \vspace{-10pt}
\begin{tabular}{|c|cccc|cccc|c|c|}
    \hline
        \multirow{2}{*}{AlexNet}   &                                                                           \multicolumn{ 8}{c|}{Design} &     {Cycles($\times$1000)}  & Elap.      \\
    \cline{2-9}
           &     $T_m$ &     $T_n$ &     $T_r$ &     $T_c$ &     $P_b$ &     $P_r$ &     $P_c$ &     $P_m$ & Comp.[+Comm.] & (sec.) \\
    \hline
    Layer1 &         96 &          3 &          1 &          55 &          4 &          1 &          1 &          1 &     375 [+290] & 0.5 \\

    Layer2 &         10 &         48 &         14 &         27 &          2 &          2 &          1 &          1 &     514 [+186] & 9.7 \\

    Layer3 &         55 &          9 &         13 &         13 &          4 &          1 &          1 &          1 &        314 [+0] & 162.3 \\

    Layer4 &         28 &         18 &         13 &         13 &          4 &          1 &          1 &          1 &    242 [+64] & 60.7\\

    Layer5 &         32 &         15 &         13 &         13 &          2 &          1 &          1 &          2 &    167 [+0] & 40.4\\
    \hline
     Total &        \multicolumn{8}{c|}{neglect reprogramming overhead}         &   2,152  & -\\
    \hline
    Cross-Layer    &       64  &   7   &   7   &   14  &  2   &   1   &   2   &   1   &   \textbf{2,239} & 797.2 \\
    \hline
\end{tabular}
\label{Tab:Cross}
\end{table}

\textbf{Multiple Layers.}
The accelerator with uniform design parameters is simple to design and can avoid the costly inter-layer communications and FPGA reconfiguration, but it may be sub-optimal for some layers.
Table \ref{Tab:Cross} gives an example of the uniform design against the layer-customized design.
From this table, we can see that the overall latency of the uniform design is 2,239 clock cycles, which is slightly (within 5\%) slower than the layer-customized design.
Kindly note that for the layer-customized design, we consider the inter-layer communication latency (in brackets), but ignore the FPGA reprogramming overhead, which will lead the layer-customized design inefficient against the uniform design.
In consequence, we use uniform design in experiments.

\revise{Finally, the column ``Elap.'' shows the elapsed time to obtain the design for layer-specific or cross-layer optimization.
As shown in this table, for the layer-specific optimization, all explorations can be finished in 3 minutes; while for the cross-layer optimization, it takes 13 minutes to obtain the design.
These results demonstrate the efficiency of the proposed formulation.}

\section{Experimental Results}\label{sec:evaluation}\label{Sec:ImpDet}

This section reports the evaluation designs synthesized by Super-LIP framework with XFER technique on Xilinx FPGAs.
Results demonstrate the significant improvements on performance and energy efficiency achieved by Super-LIP against GPUs and the existing designs.
We also demonstrate the scalability of Super-LIP and validate the accuracy of the presented system-level model.

\vspace{2pt}
\noindent\textbf{A. Implementation}
\vspace{2pt}


\begin{table*}
  \centering
  \tabcolsep 6.2pt
  \renewcommand\arraystretch{1.3}
  \footnotesize
  \caption{Experimental results of Super-LIP with comparisons to GPUs and the existing FPGA designs}
  \vspace{-10pt}
\begin{tabular}{|c|c|c|c|c|c|c|c|c|c|c|c|c|c|c|c|c|}
\hline
{\it Design}  & \multicolumn{ 2}{c|}{mGPU} & \multicolumn{ 2}{c|}{GPU}  & \multicolumn{ 2}{c|}{{\it FPGA15}} & \multicolumn{ 2}{c|}{\it ISCA17} & \multicolumn{ 2}{c|}{{\it ISLPED16}} &       \multicolumn{ 4}{c|}{{\it {\bf Super-LIP}}} \\
\hline

\hline
 Precision  & \multicolumn{ 2}{c|}{32bits float} & \multicolumn{ 2}{c|}{32bits float} & \multicolumn{ 2}{c|}{32bits float} & \multicolumn{ 2}{c|}{32bits float} & \multicolumn{ 2}{c|}{16bits fixed} & \multicolumn{ 2}{c|}{32bits float} & \multicolumn{ 2}{c|}{16bits fixed}\\
\hline
    Device  & \multicolumn{ 2}{c|}{Jetson TX2} & \multicolumn{ 2}{c|}{Titan X} & \multicolumn{ 2}{c|}{VX485T} & \multicolumn{ 2}{c|}{VX485T} & \multicolumn{ 2}{c|}{4$\times$VX690t} & \multicolumn{ 2}{c|}{2$\times$ZCU102} & \multicolumn{ 2}{c|}{2$\times$ZCU102}\\
\hline
Freq (MHz)  & \multicolumn{ 2}{c|}{1300MHz} & \multicolumn{ 2}{c|}{1139MHz} & \multicolumn{ 2}{c|}{100MHz} & \multicolumn{ 2}{c|}{100MHz} & \multicolumn{ 2}{c|}{150MHz} & \multicolumn{ 2}{c|}{100MHz} & \multicolumn{ 2}{c|}{200MHz}\\
\hline
Power (Watt) & \multicolumn{ 2}{c|}{16.00}  & \multicolumn{ 2}{c|}{162.00} & \multicolumn{ 2}{c|}{18.61} & \multicolumn{ 2}{c|}{-} & \multicolumn{ 2}{c|}{126.00} & \multicolumn{ 2}{c|}{52.40} & \multicolumn{ 2}{c|}{54.40} \\
\hline
DSP Uti.  & \multicolumn{ 2}{c|}{-} & \multicolumn{ 2}{c|}{-} & \multicolumn{ 2}{c|}{80\%} & \multicolumn{ 2}{c|}{80\%} & \multicolumn{ 2}{c|}{-} & \multicolumn{ 2}{c|}{90.79\%} & \multicolumn{ 2}{c|}{55.87\%} \\
\hline
BRAM Uti. & \multicolumn{ 2}{c|}{-} & \multicolumn{ 2}{c|}{-} & \multicolumn{ 2}{c|}{49.71\%} & \multicolumn{ 2}{c|}{43.25\%} & \multicolumn{ 2}{c|}{-} & \multicolumn{ 2}{c|}{72.92\%} & \multicolumn{ 2}{c|}{{92.43\%}} \\
\hline

\hline
\multirow{3}{*}{Overall Perf.} & Lat. & Thr. & Lat. & Thr. & Lat. & Thr. & Lat. & Thr. & Lat. & Thr. & Lat. & Thr. & Lat. & Thr. \\
                             & $ms$ & $GOPS$ & $ms$ & $GOPS$ & $ms$ & $GOPS$ & $ms$ & $GOPS$ & $ms$ & $GOPS$ & $ms$ & $GOPS$ & $ms$ & $GOPS$ \\
\cline{2-15}
    &     11.1 - 13.2 &     110.75 &      5.1 - 6.4 &    235.55&      21.62 &      69.09 &      60.13 &      85.47 &       30.6 &      128.8 & {10.13} & {149.54} & {\bf 2.27} & {\bf 679.04}  \\
\hline
E.-E. (GOPS/W) &          \multicolumn{2}{c|}{6.88} &         \multicolumn{2}{c|}{1.45}& \multicolumn{ 2}{c|}{3.71} & \multicolumn{ 2}{c|}{-} & \multicolumn{ 2}{c|}{1.02} & \multicolumn{ 2}{c|}{2.85} & \multicolumn{ 2}{c|}{{\bf 12.48}}  \\
\hline
\end{tabular}
\label{Tab:ExpComp}
\end{table*}



\textbf{System Implementation.} Figure \ref{Fig:ImpOverview} shows the implementation of a cluster with two Xilinx ZCU102 FPGAs.
FPGAs are connected by SFP+ using the Xilinx Aurora IP.
In this way, data in two FPGAs can be directly moved between their on-chip buffers.
The implementation of each FPGA utilizes the ZYNQ architecture, which controls the startup of CNN accelerator, the off-chip/on-chip communications, etc.
As shown in this figure, each FPGA has two clock domains: one for accelerator and the other for board-to-board communication.
We employ asynchronous FIFOs to coordinate data movements in different clock domains.

\begin{figure}
  \centering
  \includegraphics[width=3.3625 in]{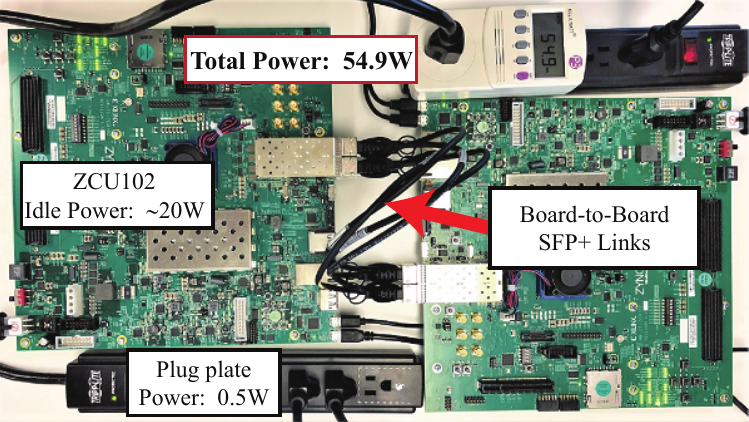}
  \vspace{-20pt}
  \caption{Power measurement of on-board executions.}\label{Fig:RealBoard}
\end{figure}

The accelerator on FPGA is implemented with Vivado HLS, which generates design's IP core from C language.
In HLS, we apply HLS-defined pragma to implement loop optimization.
Then, the obtained IP cores are connected, synthesized and implemented in Vivado (v2017.4).
In Vivado, we employ Xilinx Aurora IP core to control inter-FPGA  communication and add an axi-timer to capture the exact elapsed time.
FPGA boards are connected through SFP+ cables, as shown in Figure \ref{Fig:RealBoard}.
Finally, we employ Xilinx SDK to program MPSoC on ZCU102, which controls the start-up of the accelerator and off-chip/on-chip communication.

\textbf{Design Parameters.}
The accelerator contains two sub-systems: computation subsystem and communication subsystem.
In computation subsystem, the working frequency and computation parallelism significantly affect the performance.
We use frequency of 100MHz for floating points, while 200MHz for fixed points.
Then, the tiling parameters $\langle T_m,T_n,T_r,T_c\rangle$ determines the  computation parallelism.
These parameters can be obtained via our proposed accelerator design methodology in Super-LIP.

The communication subsystem includes the off-chip/on-chip memory communication and board-to-board communication.
The design parameters $\langle I_p,W_p,O_p\rangle$ are pre-set according to the bandwidth requirement captured from our model.
Specifically, for floating points, we set $I_p=2$, $W_p=2$, and $O_p=2$, indicating peak bandwidth is $\frac{6\times32/8}{100M}=2.4GB/s$.
For fixed points, we set $I_p=4$, $W_p=8$ and $O_p=4$, indicating peak bandwidth of $\frac{24\times16/8}{100M}=2.4GB/s$.
The data width used for board-to-board communication is set according to $I_p$ and $W_p$.
For 16bits fixed point, $W_p=8$ indicates the transmission data width is $16\times8=128$.
Kindly note that the ZCU102 board can provide the maximum data width of 256bits for bi-direction board-to-board communication.



%

In a cluster, the number of FPGAs $N$ is determined by all partitioning parameters, i.e., $N=P_b\cdot P_r\cdot P_c\cdot P_m$.
\revise{The infrastructure of the network in the FPGA cluster applies the 2D-torus topology, as illustrated in Figure \ref{Fig:Topo} of Section \ref{Sec:Hybrid}, where each FPGA has two incoming and two outgoing inter-FPGA links.}
Super-LIP will also control the data flow among FPGAs.

%
%



\vspace{2pt}
\noindent\textbf{B. Low Latency and Energy Efficiency}
\vspace{2pt}

Table \ref{Tab:ExpComp} reports the comparison results in latency, throughput and energy efficiency of AlexNet with a batch size of 1 on different platforms and designs.
The competitors of Super-LIP include mobile GPU (Jetson TX2) and GPU (Titan X), single-FPGA design (FPGA15 \cite{zhang2015optimizing}, ISCA17 \cite{shen2017maximizing}), and multi-FPGA design (ISLPED16 \cite{zhang2016energy}).
The power consumption of our implementation is measured by a power meter as demonstrated in Figure \ref{Fig:RealBoard}.
Note that notation ``-'' indicates that data is not reported in references or inapplicable.

\revise{\textbf{On-Board Measurement.} Before presenting the detailed results, we first introduce the on-board measurement, where latency and power consumption are two main metrics.
For all platforms and implementations, they will consistently process a set of  input images, say 1,000 images.
For a fair comparison, we record the latency and power consumption when the system enters the stable state (i.e., after the process of the first image).}

\revise{By recording the latency of different images, we observe that the elapsed time of GPUs is varied for different executions, while it is stable in FPGAs.
For example, the latency on mGPU ranges from 11.1ms to 13.2ms, as shown in Table \ref{Tab:ExpComp}.
It implies that the GPU implementations need to apply the worst-case execution time with an extra safety margin to satisfy the hard real-time constraint, which will drastically degrade the resource utilization and energy efficiency.}


\textbf{Latency.}
\revise{Real-time DNN inference requires ultra-low latency to avoid missing deadline.
For 32bits float-point, Super-LIP achieves latency of 10.13ms, which is 23.26\%, 2.13$\times$, 5.94$\times$ less than that of mGPU, \emph{FPGA15}, and \emph{ISCA17}.
However, Super-LIP with 32bits float-point is slower than Titan X GPU, whose latency is 6.4ms. 
This is because such GPU is much more powerful, with the penalty of consuming more than 3$\times$ power over the FPGA implementation in Super-LIP.
Benefiting from the flexibility of FPGAs to apply different data types for computation, it is possible to reduce latency by using lower-precision data type.
As shown in this table, by applying 16bits fix-point, Super-LIP can achieve the lowest latency among all competitors, i.e., 2.27ms.}








\textbf{Throughput and Energy Efficiency.}
\revise{Super-LIP makes better trade-offs between latency and throughput than the existing pipeline-based FPGA competitors (\emph{ISCA17} and \emph{ISLPED16}).
Compared with \emph{ISCA17} with 32bits float-point, Super-LIP achieves 5.94$\times$ lower latency together with 1.75$\times$ higher throughput.
The improvement in throughput is less than that on latency is because \emph{ISCA17} aims to improve throughput, but its throughput is still less than Super-LIP's.
Similarly, compare with \emph{ISLPED16} with 16bits fix-point, Super-LIP achieves 13.48$\times$ lower latency together with 5.27$\times$ higher throughput.
Benefiting from the higher throughput, Super-LIP achieves the highest energy efficiency than competitors.}



\textbf{Utilization.} From Table \ref{Tab:ExpComp}, we observe that the DSP resource is not fully utilized in Super-LIP with 16bits. This is because there is not enough BRAM resource to support parallel data access (Equations \ref{equ:dsp_flo}-\ref{equ:bram_all} for details).
By adding BRAM in the FPGA, we expect to further reduce latency and achieve higher energy efficiency. 







\begin{table}
  \centering
  \tabcolsep 4.2pt
  \renewcommand\arraystretch{1.3}
  \footnotesize
  \caption{Comparison results on ZCU102}
  \vspace{-10pt}
\begin{tabular}{|c|c|c|c|c|c|c|c|c|}
\hline
\multirow{2}{*}{Design} &               \multicolumn{ 4}{c|}{32bits float} &               \multicolumn{ 4}{c|}{16bits fixed} \\
\cline{2-9}
 & \multicolumn{ 2}{c|}{\emph{FPGA15}} & \multicolumn{ 2}{c|}{\textbf{Super-LIP}} & \multicolumn{ 2}{c|}{\emph{FPGA15}} & \multicolumn{ 2}{c|}{\textbf{Super-LIP}} \\
\hline
$\langle T_m,T_n\rangle$ & \multicolumn{ 2}{c|}{$\langle  64,7\rangle$} & \multicolumn{ 2}{c|}{$\bf \langle 64,7\rangle$} & \multicolumn{ 2}{c|}{$\langle 64,24\rangle$} & \multicolumn{ 2}{c|}{$\bf \langle 128,10\rangle$} \\
\hline
\multirow{2}{*}{Power (W)} & \multicolumn{ 2}{c|}{25.70} & \multicolumn{ 2}{c|}{\textbf{52.40}} & \multicolumn{ 2}{c|}{26.00} & \multicolumn{ 2}{c|}{\textbf{54.40}} \\
& \multicolumn{ 2}{c|}{(1 FPGA)} & \multicolumn{ 2}{c|}{\textbf{(2 FPGAs)}} & \multicolumn{ 2}{c|}{(1 FPGA)} & \multicolumn{ 2}{c|}{\textbf{(2 FPGAs)}} \\
\hline
\hline
\multirow{ 2}{*}{Perf.} &       Lat. &       Thr. &      \textbf{ Lat.} &       \textbf{Thr.} &       Lat. &       Thr. &       \textbf{Lat.} &       \textbf{Thr.} \\
           &         ms &       GOPS &         \textbf{ms} &       \textbf{GOPS} &         ms &       GOPS &         \textbf{ms} &       \textbf{GOPS} \\
\hline
     conv1 &       7.36 &      28.6 &       \textbf{3.66} &      \textbf{57.6 }&        3.74    &    56.5        &       \textbf{0.94} &     \textbf{224.5} \\
\hline
     conv2 &       5.20 &      86.1 &       \textbf{2.55} &     \textbf{175.5} &        1.48    &    302.6        &       \textbf{0.48} &     \textbf{933.1} \\
\hline
     conv3 &       4.50 &      66.4 &       \textbf{1.73} &     \textbf{172.7} &        1.20    &    249.6        &       \textbf{0.33} &     \textbf{906.2} \\
\hline
     conv4 &       3.41 &      65.7 &       \textbf{1.31} &     \textbf{171.0} &        0.89    &    252.6        &       \textbf{0.35} &     \textbf{640.8} \\
\hline
     conv5 &       2.28 &      66.0 &       \textbf{0.88} &     \textbf{170.9} &        0.59    &    251.7        &       \textbf{0.17} &     \textbf{879.5} \\
\hline
   overall &      22.75 &      66.6 &      \textbf{10.13} &     \textbf{149.5} &        7.90    &    195.1        &       \textbf{2.27} &     \textbf{679.0} \\
\hline
Perf. Impr. & \multicolumn{ 2}{c|}{1.00$\times$} & \multicolumn{ 2}{c|}{$\bf 2.25 \times$} & \multicolumn{ 2}{c|}{1.00$\times$} &  \multicolumn{ 2}{c|}{\textbf{3.48}$\bf \times$} \\
\hline
\hline
E.-E. & \multicolumn{2}{c|}{\multirow{2}{*}{2.59}} & \multicolumn{ 2}{c|}{\multirow{2}{*}{\textbf{2.85}}} &  \multicolumn{ 2}{c|}{\multirow{2}{*}{7.51}} & \multicolumn{ 2}{c|}{\multirow{2}{*}{\textbf{12.48}}} \\
(GOPS/W) & 		\multicolumn{ 2}{c|}{}	&		\multicolumn{ 2}{c|}{}	&		\multicolumn{ 2}{c|}{}	&	\multicolumn{ 2}{c|}{}		\\
\hline
E.-E. Impr. & \multicolumn{ 2}{c|}{-} & \multicolumn{ 2}{c|}{\textbf{9.21\%}} & \multicolumn{ 2}{c|}{-} &  \multicolumn{ 2}{c|}{\textbf{39.86\%}} \\
\hline
\end{tabular}  
\label{Tab:ZCUComp}
\end{table}

\begin{table*}
  \centering
  \tabcolsep 5.4pt
  \renewcommand\arraystretch{1.3}
  \footnotesize
  \caption{\revise{Experimental results on model validation, performance bottleneck detection and alleviation}}
  \vspace{-10pt}
\begin{tabular}{|c|c|c|c|c|c|c|c|c|c|c|c|c|c|c|}
\hline
\multirow{2}{*}{Design} & \multirow{2}{*}{Precision} & \multirow{2}{*}{$\langle Tm,Tn \rangle$ } & \multirow{2}{*}{Partition} &     \multicolumn{ 4}{c|}{Our Model}             &      \multicolumn{ 3}{c|}{On-Board} & \multicolumn{ 3}{c|}{{\bf Deviation}} & \multirow{2}{*}{{\bf Speedup}} \\
\cline{5-14}
 &  &  &  &     Cycles &       BRAM &       DSPs & \textbf{Bound} &     Cycles &       BRAM &       DSPs & {\bf Cycles} & {\bf BRAM} & {\bf DSPs} & \multicolumn{ 1}{c|}{{\bf }} \\
\hline
         A (Single) & \multirow{2}{*}{32b float} &     $\langle8,32 \rangle$ &          - &     519168 &        592 &       1280 &        \textbf{IFM} &     535530 &        624 &       1326 & {\bf 3.06\%} & {\bf 5.13\%} & {\bf 3.47\%} &   {\bf baseline} \\
\cline{1-1}\cline{3-15}
         B (XFER) &  &     $\langle8,32 \rangle$ &       Pm=2 &     158880 &        592 &       1280 &      \textbf{Comp.} &     162114 &        640 &       1331 & {\bf 1.99\%} & {\bf 7.50\%} & {\bf 3.83\%} & {\bf 3.30X} \\
\hline
         C (Single)& \multirow{2}{*}{16b fixed} &    $\langle64,20 \rangle$ &          - &     115200 &       1448 &       1280 &     \textbf{Weight} &     118688 &       1516 &       1324 & {\bf 2.94\%} & {\bf 4.49\%} & {\bf 3.32\%} &   {\bf baseline} \\
\cline{1-1}\cline{3-15}
         D (XFER) &  &    $\langle64,20 \rangle$ &       Pr=2 &      32760 &       1448 &       1280 &      \textbf{Comp.} &      34622 &       1530 &       1330 & {\bf 5.38\%} & {\bf 5.36\%} & {\bf 3.76\%} & {\bf 3.43X} \\
\hline
\end{tabular}
\label{Tab:ExpModel}
\end{table*}

\vspace{2pt}
\noindent\textbf{C. Results on ZCU102}
\vspace{2pt}

We notice that, in Table \ref{Tab:ExpComp}, the energy efficiency of Super-LIP with 32bits is less than \emph{FPGA15}. This is because the power consumed by ZCU102 in idle state (i.e., $\sim20W$) is already larger than that of \emph{FPGA15} at run-time (\emph{18.61W}).
For a fair comparison, we re-implement \emph{FPGA15} on ZCU102, and the results are reported in Table \ref{Tab:ZCUComp}.



\textbf{32bits float-point.} We have several conclusions. First, the design parameters ($T_m$ and $T_n$) of \emph{FPGA15} and Super-LIP are the same; therefore, the only difference is that Super-LIP has extra inter-FPGA communication.
Second, the power in \emph{FPGA15} design (25.70W) is less than half of that in Super-LIP (52.40W). 
The gap $52.40-25.70\times 2=1.0W$ on power consumption is caused by the inter-FPGA communication (including IP core, communication links, etc.), which only occupies 1.91\% of the total power.
Third, Super-LIP with 2 FPGAs achieves $2.25\times$ speedup over \emph{FPGA15}, indicating that we have achieved super-linear speedup.
Furthermore, benefiting from the significant performance achievement, Super-LIP obtains 9.21\% improvement in energy efficiency.

\textbf{16bits fix-point.} Unlike 32bits float-point, the optimal design parameters ($\langle T_m,T_n \rangle$) explored by \emph{FPGA15} and Super-LIP are different.
This is because the design $\langle 128,10 \rangle$ alleviates bottleneck on memory bandwidth, which results in severe performance degradation in the overall assessment for \emph{FPGA15} design, while Super-LIP can resolve such bottlenecks to achieve better performance.
Specifically, our Super-LIP design has achieved 3.48$\times$ speedup and 39.86\% improvements in energy efficiency, when compared with \emph{FPGA15}.

\vspace{2pt}
\noindent\textbf{\revise{D. Model Accuracy and Effectiveness}}
\vspace{2pt}

\begin{figure}
  \centering
  \includegraphics[width=2.9in]{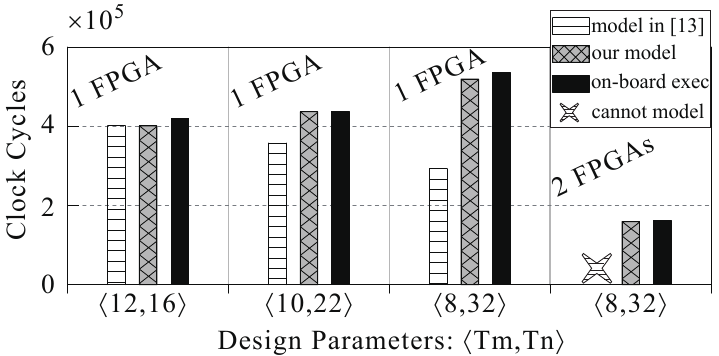}
  \vspace{-10pt}
  \caption{\revise{Comparisons of predictable models and on-board executions on latency: employing different designs on single-FPGA and 2-FPGA systems.}}\label{Fig:ModelComp}
\end{figure}

\revise{Now, we are going to validate the accuracy and effectiveness of the proposed system-level model.
We will conduct two sets of experiments: (1) we compare the proposed model with the existing one in predict latency; (2) we compare the proposed model with the final implementation results from Vivado in memory resource, computation resource, and on-board execution latency.}

\revise{First, Figure \ref{Fig:ModelComp} reports the comparison results among different models and on-board execution latency.
The x-axis and y-axis represent different designs and latency in clock cycles, respectively.
In the first three designs, we employ one FPGA for implementation; while for the fourth one, we employ 2 FPGAs.}

\revise{Results in Figure \ref{Fig:ModelComp} clearly show that the latency predicted by our proposed model is always close to the on-board execution latency, where the average deviation is only 2.53\%.
In contrast, the existing model in \cite{zhang2015optimizing} has larger deviations on designs of $\langle 10,22\rangle$ and $\langle 8,32\rangle$, which are 18.49\% and 45.47\%.
In addition, the existing model cannot predict for multiple FPGAs, but ours can.}

\revise{We have another observation from Figure \ref{Fig:ModelComp}. For the design of $\langle 12,16\rangle$, model in \cite{zhang2015optimizing} predicts the same latency with ours. 
This is because the computation latency dominates the whole system.
In this case, the inaccurate estimation of communication will not affect prediction accuracy.
However, when we employ more computation resource (by increasing $T_m\times T_n$), the performance bottleneck moves to communication which leads to the large latency deviations between the existing model and the on-board execution.}

\revise{The above results verify the accuracy and effectiveness of the proposed system-level model in predicting system latency.
With such an accurate model, it can help designers to get the accurate system performance to make better design decisions.}

\revise{Next, Table \ref{Tab:ExpModel} reports the comparison between the proposed model and the final implementation results from Vivado in BRAMs and DSPs. It is clear that the deviations on BRAM and DSP usages are less than 7.5\% and 3.9\%, respectively.
These deviations are mainly caused by the overhead on extra operations besides the accelerator itself, such as DSPs used for address calculation. 
The above results further verify the accuracy of the proposed model.}





\revise{Results in Table \ref{Tab:ExpModel} further verify the effectiveness of the proposed techniques in detecting and alleviating system performance bottlenecks. 
For the single-FPGA designs A and C, we employ Corollary \ref{Corollary1}  to detect their performance bottlenecks, as shown in Column ``Bound'' under Column ``Our Model''.
It indicates that the performance of design A is bounded by loading IFM data, while that of design C is bounded by loading weights data.
For design A, we apply XFER technique by setting $P_m=2$ to share IFM data on inter-FPGA links, and it outputs the design B.
As shown in this table, the performance bottleneck on design B has been successfully moved to computation, and therefore, it achieves 3.3$\times$ speedup.
Similarly, we apply XFER technique on design C to generate design D with 3.43$\times$ speedup.}

\revise{The above results validate the accuracy of the proposed model in modeling system resources.
In addition, the proposed system-level model can be applied to effectively detect performance bottleneck.
And the proposed XFER technique can be applied to  alleviate different kinds of bottlenecks.
As a result, Super-LIP can achieve super-linear speedups with multiple FPGAs.}





\vspace{2pt}
\noindent\textbf{E. Design Space Exploration}
\vspace{2pt}

\begin{figure}
  \centering
  \includegraphics[width=3.2096 in]{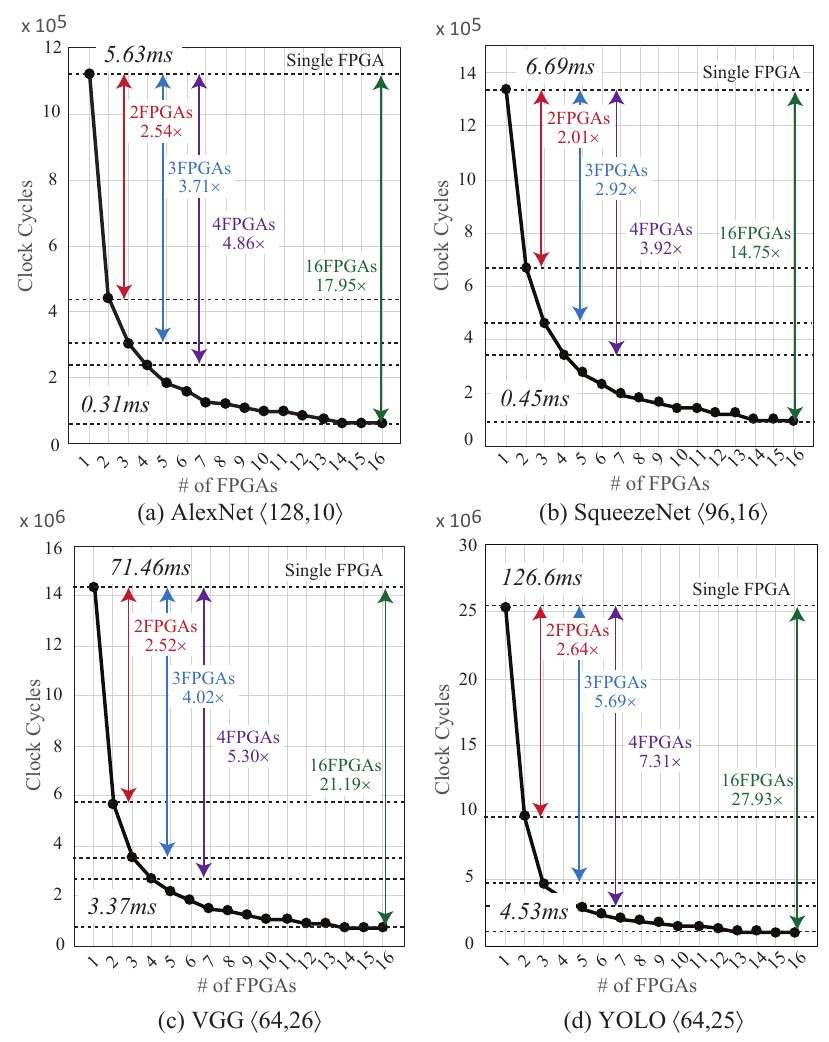}
  \vspace{-6pt}
  \caption{Design space exploration of Super-LIP with the increasing number of FPGAs using different CNNs.}\label{Fig:ExpScala}
\end{figure}

Finally, we explore the design space to demonstrate the scalability of Super-LIP.
Specifically, we scale up the number of FPGAs with the same design parameters (the optimal ones in single-FPGA design) but different partitions.
Figure \ref{Fig:ExpScala} reports the experimental results of four widely used CNNs with 16bits fixed point on the clusters with up to 16 FPGAs, including AlexNet, SqueezeNet, VGG, and YOLO.
In the figure, the x-axis and y-axis represent the number of involved FPGAs and the total clock cycles.
Each point corresponds to a design with specific loop tiling and partition factors.
We give the tiling value ($T_m$ and $T_n$) of CNNs in each sub-figure; for instance, in AlexNet, $T_m=128$ and $T_n=10$.
\revise{For the system with no more than 2 FPGAs, we have implemented the accelerators on the testbed in Figure \ref{Fig:RealBoard}, and obtain the on-board execution latency.
While for larger FPGA cluster, we obtain the latency using the following method.
First, according to these tiling factors, we implement the accelerator on FPGA to obtain its on-board execution latency.
Then, according to the partition factors for each design point, we can obtain how many times the accelerator will be invoked in each layer.
Based on the above two kinds of information, we can get the computation latency.
In addition, according to the determined partition factors, we can get the communication load on intra-FPGA and inter-FPGA links to calculate the communication latency.
Finally, the overall latency can be derived from Formula \ref{equ:Lat}.}

As shown in Figures \ref{Fig:ExpScala}(a)-(d), with the increasing number of FPGAs, Super-LIP can consistently reduce the overall latency.
We observe that the speedup of SqueezeNet is relatively small, mainly because the sizes of weight and IFM are small owing to the squeeze operations.
However, its latency can still be drastically reduced, from \emph{6.69ms} to \emph{0.45ms}.
For AlexNet, VGG, and YOLO, super-linear performance can be consistently achieved with 2-16 FPGAs.
Specifically, for YOLO, the latency is reduced from \emph{126.6ms} to \emph{4.53ms} with 16 FPGAs, achieving $27.93\times$ speedup.

\revise{We have another observation from Figure \ref{Fig:ExpScala}(b): for SqueezeNet, when the number of FPGAs increases to 3, the speedup is only 3.92$\times$, which indicates the failure of achieving super-linear speedup.
This is because SqueezeNet contains many convolution operations with the kernel size of 1, which leads the performance bottleneck mainly on computation.  
In contrast, we can consistently achieve super-linear speedup for the other three CNNs when the size of FPGA cluster scales up to 16.
However, when the size of FPGA cluster scales up, the linear performance will be terminated since the number of channels (or row, column) in a CNN layer is fixed, and we cannot further improve parallelism by adding  partition factor when it reaches the maximum number.}

\revise{Benefiting from the super-linear performance, the system energy efficiency can be improved. 
Compared with the single FPGA design, for AlexNet, VGG, and YOLO on 4 FPGAs, the energy efficiency improvements are 11.29\%, 20.65\%, and 41.02\%, respectively.
When the size of FPGA cluster scales up to 16, these figures are 3.93\%, 18.61\%, and 36.25\%.
We observe that the improvements are decreased because the overheads on communication are increased as the cluster size increases.
Overall, these results verify that by applying the proposed techniques on multiple FPPGAs, we can achieve super-linear speedup, and in turn, the energy efficiency can be improved against the single-FPGA design.}

Finally, we analyze the bandwidth requirement of clusters.
From Figures \ref{Fig:ExpScala}(a)-(d), we can observe that the performance improvement is converged when the number of FPGAs approaches 16 with a $4\times 4$ torus topology.
In this scale, the inter-FPGA bandwidth provided by ZCU102 is sufficient.
Specifically, for each FPGA, there are 3 weights and 3 IFMs needing to be transmitted simultaneously. 
Thus, the BW requirement for each FPGA is $(3+3)\times(16bits/cycle)=144bits/cycle$; while ZCU102 provides a bandwidth of 256bits/cycle (4 SPF+ ports with 64bits wide each). 
Furthermore, we can add 4 QSFP ports for additional bandwidth of $4\times 256=1024bits/cycle$ for even larger clusters.

The above results verify the scalability of Super-LIP.
In addition, our techniques are effective to explore the design space to provide more options for different timing constraints.

\section{Related Work}




The development of FPGA-based DNNs accelerator evolves in three stages.
At the early stage \cite{ouyang2014sda,zhang2015optimizing,venieris2016fpgaconvnet,xu2018scaling,xu2018resource,xu2018quantization}, the whole FPGA is designed as one accelerator, and a controller iteratively moves data from off-chip DRAM to the accelerator to be executed. 
In the second stage, it is observed that the computation resource cannot be fully utilized with the one-size accelerator due to the varied computation and memory requirements in DNN layers.
To overcome this shortage, multiple accelerators are integrated into one FPGA \cite{guo2016angel,shen2017maximizing,zhang2018dnnbuilder}.
However, the restrict resource on one board still limits the performance boosting of DNNs on FPGAs.

\revise{Most recently, with the growing demand in time performance, it is a trend to employ a cluster of FPGAs to execute DNNs \cite{zhang2016energy,jiang2018heterogeneous,geng2018fpdeep,zhang2019efficient,shen2019scale,shen2019accelerating,jiang2019accuracy,jiang2019hardware}.
In \cite{zhang2016energy,zhang2019efficient}, authors construct multiple FPGAs as a pipeline to execute a set of input images in a pipeline fashion.
In \cite{jiang2018heterogeneous}, authors split the CNN layers to balance pipeline stages for higher throughput and lower cost.
Authors in \cite{geng2018fpdeep} employ multiple FPGAs for the training phase.
In \cite{shen2019scale,shen2019accelerating}, multi-FPGA platforms are utilized to accelerate the lung nodule segmentation.
All the above works target on improving throughput by using a pipeline of FPGAs, which can achieve high throughput but make sacrifices on latency.}

To satisfy the low latency requirement for real-time DNN inference, Microsoft in Brainwave \cite{chung2018serving,fowers2018configurable} devise techniques to pin weights on different FPGAs.
Such an approach can work well for RNNs with small intermediate data, but awkward for CNN implementations due to the large intermediate data and complicated data reuse pattern.
Kindly note that in \cite{chung2018serving,fowers2018configurable}, authors use only one FPGA for CNNs, whose input image has low resolution that hides the bandwidth bottleneck issue. 
However, for more realistic CNN applications with high resolution, like medical images, it is still unknown how to achieve real-time inference with ultra-low latency using multiple FPGAs.
Super-LIP is proposed to fill this gap.

\revise{Another branch of related work is to deploy CNNs on multi-core mobile devices or multi-processor system on-chip (MPSoC) \cite{wang2018optic,motamedi2018cappuccino,yang2018optimal,wang2018towards,wang2018exploiting,wu2019towards}.
Unlike FPGA-based implementation that requires designers to determine the designs of communication and computation sub-systems, processing elements in these systems use fixed designs (e.g., CPUs, GPUs).
In consequence, the optimization problem on such systems is how to run tasks to computation components in parallel, without considering how to tailor hardware designs.}

\section{\revise{Conclusion and Future Work}}

In this work, we propose the Super-LIP framework to achieve super-linear speedup for Deep Neural Networks (DNNs) inference on multi-FPGA cluster.
We formulate an accuracy model to design accelerators and matched performance bottleneck detection techniques.
In addition, we propose XFER design, a novel design for multi-FPGA cluster to minimize the overall system latency without compromising throughput or energy efficiency, such that the resultant system can provide real-time inference.
As a case study, we implement CNN on a small-scale FPGA cluster with two Xilinx ZCU102 boards connected via SFP+.
Evaluation results show that the proposed Super-LIP framework with 2 FPGAs can achieve $3.48\times$ speedup compared with the FPGA design in \cite{zhang2015optimizing}, meanwhile, achieving $39.86\%$ improvement on energy efficiency.

\revise{In terms of the rapid development of computing infrastructure in both edges and clouds, the platform is evolving to compose heterogeneous (different types) FPGAs.
Kindly note that the accurate models and the XFER design will be the base for the cluster with heterogeneous FPGAs.
In the future, we will develop optimal algorithms to optimize latency in the heterogeneous platforms.}





\begin{thebibliography}{10}

\bibitem{krizhevsky2012imagenet}
A.~Krizhevsky {\em et~al.}, ``Imagenet classification with deep convolutional
  neural networks,'' in {\em Proc. of NIPS}, pp.~1097--1105, 2012.

\bibitem{ren2017faster}
S.~Ren {\em et~al.}, ``Faster r-cnn: towards real-time object detection with
  region proposal networks,'' {\em IEEE TPAMI}, no.~6, pp.~1137--1149, 2017.

\bibitem{young2017recent}
T.~Young {\em et~al.}, ``Recent trends in deep learning based natural language
  processing,'' {\em arXiv preprint arXiv:1708.02709}, 2017.

\bibitem{zhang2017live}
H.~Zhang {\em et~al.}, ``Live video analytics at scale with approximation and
  delay-tolerance.,'' in {\em Proc. of NSDI}, vol.~9, p.~1, 2017.

\bibitem{balakrishnan2018unsupervised}
G.~Balakrishnan {\em et~al.}, ``An unsupervised learning model for deformable
  medical image registration,'' in {\em Proc. of CVPR}, pp.~9252--9260, 2018.

\bibitem{chung2018serving}
E.~Chung {\em et~al.}, ``Serving dnns in real time at datacenter scale with
  project brainwave,'' {\em IEEE Micro}, vol.~38, no.~2, pp.~8--20, 2018.

\bibitem{fowers2018configurable}
J.~Fowers {\em et~al.}, ``A configurable cloud-scale dnn processor for
  real-time ai,'' in {\em Proc. of ISCA}, pp.~1--14, IEEE Press, 2018.

\bibitem{ding2018universal}
Y.~Ding {\em et~al.}, ``On the universal approximability and complexity bounds
  of quantized relu neural networks,'' {\em arXiv preprint arXiv:1802.03646},
  2018.

\bibitem{wilhelm2008worst}
R.~Wilhelm {\em et~al.}, ``The worst-case execution-time problem-overview of
  methods and survey of tools,'' {\em ACM TECS}, vol.~7, no.~3, p.~36, 2008.

\bibitem{jouppi2017datacenter}
N.~P. Jouppi {\em et~al.}, ``In-datacenter performance analysis of a tensor
  processing unit,'' in {\em Proc. of ISCA}, pp.~1--12, IEEE, 2017.

\bibitem{ma2017optimizing}
Y.~Ma {\em et~al.}, ``Optimizing loop operation and dataflow in fpga
  acceleration of deep convolutional neural networks,'' in {\em Proc. of FPGA},
  pp.~45--54, ACM, 2017.

\bibitem{shen2017maximizing}
Y.~Shen {\em et~al.}, ``Maximizing cnn accelerator efficiency through resource
  partitioning,'' in {\em Proc. of ISCA}, pp.~535--547, IEEE, 2017.

\bibitem{suda2016throughput}
N.~Suda {\em et~al.}, ``Throughput-optimized opencl-based fpga accelerator for
  large-scale convolutional neural networks,'' in {\em Prof. of FPGA},
  pp.~16--25, ACM, 2016.

\bibitem{zhang2015optimizing}
C.~Zhang {\em et~al.}, ``Optimizing fpga-based accelerator design for deep
  convolutional neural networks,'' in {\em Proc. of FPGA}, pp.~161--170, ACM,
  2015.

\bibitem{zhang2016energy}
C.~Zhang {\em et~al.}, ``Energy-efficient cnn implementation on a deeply
  pipelined fpga cluster,'' in {\em Proc. of ISLPED}, pp.~326--331, ACM, 2016.

\bibitem{redmon2016you}
J.~Redmon {\em et~al.}, ``You only look once: Unified, real-time object
  detection,'' in {\em Proc. of CVPR}, pp.~779--788, 2016.

\bibitem{yang2017task}
L.~Yang {\em et~al.}, ``Task mapping on smart noc: Contention matters, not the
  distance,'' in {\em Proc. DAC}, pp.~1--6, IEEE, 2017.

\bibitem{yang2016fotonoc}
L.~Yang {\em et~al.}, ``Fotonoc: A folded torus-like network-on-chip based
  many-core systems-on-chip in the dark silicon era,'' {\em IEEE TPDS},
  vol.~28, no.~7, pp.~1905--1918, 2016.

\bibitem{ouyang2014sda}
J.~Ouyang {\em et~al.}, ``Sda: Software-defined accelerator for large-scale dnn
  systems,'' in {\em Proc. of HCS}, pp.~1--23, IEEE, 2014.

\bibitem{venieris2016fpgaconvnet}
S.~I. Venieris {\em et~al.}, ``fpgaconvnet: A framework for mapping
  convolutional neural networks on fpgas,'' in {\em Proc. of FCCM}, pp.~40--47,
  IEEE, 2016.

\bibitem{xu2018scaling}
X.~Xu {\em et~al.}, ``Scaling for edge inference of deep neural networks,''
  {\em Nature Electronics}, vol.~1, no.~4, p.~216, 2018.

\bibitem{xu2018resource}
X.~Xu {\em et~al.}, ``Resource constrained cellular neural networks for
  real-time obstacle detection using fpgas,'' in {\em Proc. ISQED},
  pp.~437--440, IEEE, 2018.

\bibitem{xu2018quantization}
X.~Xu {\em et~al.}, ``Quantization of fully convolutional networks for accurate
  biomedical image segmentation,'' in {\em Proc. CVPR}, pp.~8300--8308, 2018.

\bibitem{guo2016angel}
K.~Guo {\em et~al.}, ``Angel-eye: A complete design flow for mapping cnn onto
  customized hardware,'' in {\em Proc. of ISVLSI}, pp.~24--29, IEEE, 2016.

\bibitem{zhang2018dnnbuilder}
X.~Zhang {\em et~al.}, ``Dnnbuilder: an automated tool for building
  high-performance dnn hardware accelerators for fpgas,'' in {\em Proc. of
  ICCAD}, p.~56, ACM, 2018.

\bibitem{jiang2018heterogeneous}
W.~Jiang {\em et~al.}, ``Heterogeneous fpga-based cost-optimal design for
  timing-constrained cnns,'' {\em IEEE TCAD}, vol.~37, no.~11, pp.~2542--2554,
  2018.

\bibitem{geng2018fpdeep}
T.~Geng {\em et~al.}, ``Fpdeep: Acceleration and load balancing of cnn training
  on fpga clusters,'' in {\em Proc. of FCCM}, pp.~81--84, IEEE, 2018.

\bibitem{zhang2019efficient}
W.~Zhang {\em et~al.}, ``\revise{An Efficient Mapping Approach to Large-Scale
  DNNs on Multi-FPGA Architectures},'' in {\em Proc. DATE}, pp.~1241--1244,
  IEEE, 2019.

\bibitem{shen2019scale}
J.~Shen {\em et~al.}, ``\revise{Scale-out Acceleration for 3D CNN-based Lung
  Nodule Segmentation on a Multi-FPGA System},'' in {\em Proc. DAC}, p.~207,
  ACM, 2019.

\bibitem{shen2019accelerating}
J.~Shen {\em et~al.}, ``\revise{Accelerating 3D CNN-based Lung Nodule
  Segmentation on a Multi-FPGA System},'' in {\em Proc. FPGA}, pp.~117--117,
  ACM, 2019.

\bibitem{jiang2019accuracy}
W.~Jiang {\em et~al.}, ``Accuracy vs. efficiency: Achieving both through
  fpga-implementation aware neural architecture search,'' in {\em Proc. DAC},
  p.~5, ACM, 2019.

\bibitem{jiang2019hardware}
W.~Jiang {\em et~al.}, ``Hardware/software co-exploration of neural
  architectures,'' {\em arXiv preprint arXiv:1907.04650}, 2019.

\bibitem{wang2018optic}
S.~Wang {\em et~al.}, ``\revise{OPTiC: Optimizing Collaborative CPU--GPU
  Computing on Mobile Devices With Thermal Constraints},'' {\em IEEE TCAD},
  vol.~38, no.~3, pp.~393--406, 2018.

\bibitem{motamedi2018cappuccino}
M.~Motamedi {\em et~al.}, ``\revise{Cappuccino: Efficient cnn inference
  software synthesis for mobile system-on-chips},'' {\em IEEE ESL}, vol.~11,
  no.~1, pp.~9--12, 2018.

\bibitem{yang2018optimal}
L.~Yang {\em et~al.}, ``Optimal application mapping and scheduling for
  network-on-chips with computation in stt-ram based router,'' {\em IEEE
  Transactions on Computers}, 2018.

\bibitem{wang2018towards}
Y.~Wang {\em et~al.}, ``Towards memory-efficient allocation of cnns on
  processing-in-memory architecture,'' {\em IEEE TPDS}, vol.~29, no.~6,
  pp.~1428--1441, 2018.

\bibitem{wang2018exploiting}
Y.~Wang {\em et~al.}, ``Exploiting parallelism for cnn applications on 3d
  stacked processing-in-memory architecture,'' {\em IEEE TPDS}, vol.~30, no.~3,
  pp.~589--600, 2018.

\bibitem{wu2019towards}
S.~Wu {\em et~al.}, ``Towards cross-platform inference on edge devices with
  emerging neuromorphic architecture,'' in {\em Proc. DATE}, pp.~806--811,
  IEEE, 2019.

\end{thebibliography}
\end{document}